\documentclass[
reprint,
superscriptaddress,
amsmath,
amssymb,
showkeys,
aps,
prb,
floatfix,
]{revtex4-2}

\usepackage{threeparttable}
\usepackage{setspace}
\usepackage{makecell}
\usepackage{amsmath}
\usepackage{graphicx}
\usepackage{dcolumn}
\usepackage{bm}
\usepackage{hyperref}
\usepackage[mathlines, pagewise]{lineno}
\usepackage{textcomp}
\usepackage{gensymb}
\usepackage{subfigure}
\usepackage{url}
\usepackage{xcolor}
\usepackage{soul}
\usepackage{hhline}
\usepackage[version=3]{mhchem}
\usepackage{multirow}
\usepackage[american]{babel}

\begin{document}
%\linenumbers

\title{Complex \texorpdfstring{\ce{Ga2O3}}{} Polymorphs Explored by Accurate and General-Purpose Machine-Learning Interatomic Potentials}

\author{Junlei Zhao} \email{zhaojl@sustech.edu.cn}
\affiliation{Department of Electrical and Electronic Engineering, Southern University of Science and Technology, Shenzhen 518055, China}

\author{Jesper Byggm{\"a}star} %\email{jesper.byggmastar@helsinki.fi}
\affiliation{Department of Physics, University of Helsinki, P.O. Box 43, FI-00014, Finland}
\affiliation{FCAI: Finnish Center for Artificial Intelligence, University of Helsinki, P.O. Box 43, FI-00014, Finland}

\author{Huan He} %\email{he.huan@helsinki.fi}
\affiliation{Department of Physics, University of Helsinki, P.O. Box 43, FI-00014, Finland}
\affiliation{School of Nuclear Science and Technology, Xi'an Jiaotong University, Xi'an, Shaanxi 710049, China}

\author{Kai Nordlund} %\email{kai.nordlund@helsinki.fi}
\affiliation{Department of Physics, University of Helsinki, P.O. Box 43, FI-00014, Finland}
\affiliation{Helsinki Institute of Physics, University of Helsinki, P.O. Box 43, FI-00014, Finland}

\author{Flyura Djurabekova} %\email{flyura.djurabekova@helsinki.fi}
\affiliation{Department of Physics, University of Helsinki, P.O. Box 43, FI-00014, Finland}
\affiliation{Helsinki Institute of Physics, University of Helsinki, P.O. Box 43, FI-00014, Finland}

\author{Mengyuan Hua} \email{huamy@sustech.edu.cn}
\affiliation{Department of Electrical and Electronic Engineering, Southern University of Science and Technology, Shenzhen 518055, China}


\begin{abstract}

\ce{Ga2O3} is a wide-bandgap semiconductor of emergent importance for applications in electronics and optoelectronics. 
However, vital information of the properties of complex coexisting \ce{Ga2O3} polymorphs and low-symmetry disordered structures is missing. 
In this work, we develop two types of kernel-based machine-learning Gaussian approximation potentials (ML-GAPs) for \ce{Ga2O3} with high accuracy for $\beta$/$\kappa$/$\alpha$/$\delta$/$\gamma$ polymorphs and generality for disordered stoichiometric structures. 
We release two versions of interatomic potentials in parallel, namely soapGAP and tabGAP, for excellent accuracy and exceeding speedup, respectively. 
We systematically show that both the soapGAP and tabGAP can reproduce the structural properties of all the five polymorphs in an exceptional agreement with \textit{ab initio} results, meanwhile boost the computational efficiency with $5\times10^{2}$ and $2\times10^{5}$ computing speed increases compared to density functional theory, respectively. 
The results show that the liquid-solid phase transition proceeds in three different stages, a ``slow transition", ``fast transition", and ``only \ce{Ga} migration". 
We further show that this experimentally unrevealed complex dynamics can be understood in terms of distinctly different mobilities of \ce{O} and \ce{Ga} sublattices in the interfacial layer.

\end{abstract}

\maketitle

\section{Introduction} 

\ce{Ga2O3} emerges as a vital next-generation wide-bandgap semiconductor that has stimulated enormous research and application interests over the last decade~\cite{pearton2018a, jesenovec2022persistent}. 
Desirable physical properties, such as a wide bandgap ($4.8\sim5.3$ eV), a large critical electric field ($\sim8$ MV/cm), a small electron effective mass ($0.27\sim0.28m_{0}$), and transparency deep into the ultraviolet (UV) range, bring \ce{Ga2O3} forward for highly promising utility, especially in the field of high-power electronics~\cite{pearton2018perspective, zhang2022ultra}, solar-blind UV detectors~\cite{kim2020highly, wang2021ultrahigh, tang2022quasiepitaxial}, high-temperature gas sensing~\cite{mazeina2010functionalized, SFzhao2021two}, and low-dimensional devices~\cite{zavabeti2017liquid, wurdack2021ultrathin, SFzhao2022bilayer}.

A particular difficulty for precise engineering of \ce{Ga2O3} largely arises from the complex polymorphic nature of this material. 
Five different polymorphs of \ce{Ga2O3}, namely $\alpha$, $\beta$, $\gamma$, $\delta$ and $\kappa$, are analogous to those of \ce{Al2O3}. 
In addition, there are somewhat inconsistent descriptions of a possible $\epsilon$-phase~\cite{playford2013structures, cora2017real, swallow2020influence, mu2022phase}.
The latest experimental finding~\cite{cora2017real} indicates that the $\epsilon$-phase consists of three $\kappa$-phase domains connected with 120$\degree$ pseudo-hexagonal symmetry. 
Therefore, for consistency, we will refer to this phase herein as the $\kappa$-phase of the space group $Pna2_{1}$.

The most stable \ce{Ga2O3} phase at room temperature and ambient pressure is the monoclinic $\beta$-phase. 
Some metastable phases, such as the $\kappa$- and $\alpha$-phases, -- the second and the third most stable phases, respectively -- can be synthesized under high-temperature/pressure conditions~\cite{remeika1966growth, lion2022elastic}, or by means of carefully controlled thin-film growth methods~\cite{sun2018hcl, wheeler2020phase}.
Moreover, some properties of \ce{Ga2O3} in these metastable phases were reported to be superior to those of the $\beta$-phase.
For example, the hexagonal $\alpha$-phase has a still wider bandgap ($5.25\sim5.3$ eV~\cite{he2006first}) than the $\beta$-phase. 
The $\alpha$-phase can grow heteroepitaxially on a $c$-plane of sapphire substrates with higher quality than that grown from the $\beta$-phase~\cite{uno2020growth}. 
In turn, the $\kappa$-phase has demonstrated ferroelectric properties~\cite{mezzadri2016crystal, kim2018first, ranga2020highly} favorable for generation of high-density two-dimensional electron gas (2DEG) in power devices.
Therefore, the precise control over \ce{Ga2O3} polymorphs (phase engineering) is of great research interest and under active experimental scrutiny. 
Formation of strain-induced metastable phases were also observed not only during low-temperature
heteroepitaxial growth on sapphire substrates~\cite{xu2019strain, cora2020insitu}, but also under high fluence strongly focused ion-beam radiation~\cite{tetelbaum2021ionbeam, azarov2022disorder}.
However, subsequent annealing may cause strain relief triggering solid-state phase transition back to the most stable $\beta$-phase~\cite{zhu2016mechanism, cora2020insitu}.

Extensive experimental studies of \ce{Ga2O3} this far were complemented only by computational modeling on small-scale \ce{Ga2O3} atomic systems (less than 10$^{3}$ atoms) using computationally expensive \textit{ab initio} calculations.  
Recent results obtained by density functional theory (DFT) for \ce{Ga2O3} report accurate structural and electronic properties of the perfect bulk phases~\cite{schubert2016anisotropy, furthm2016quasi, ponce2020structural, mu2022phase}, formation energies of point defects~\cite{ varley2016oxygenerr, peelaers2019deep}, and surfaces~\cite{peelaers2017lack, mu2020first}. 
However, to study \ce{Ga2O3} phase engineering that involves solid-state phase transition in thousand- or tens-of-thousand-atom systems with coexisting polymorphs, one will inevitably need large-scale classical atomistic modelling using molecular dynamics (MD) methods.
Carrying out MD simulations has been hindered by the high complexity of the \ce{Ga2O3} polymorphs and the lack of reliable interatomic potential (force field) models.
Fortunately, accurate large-scale MD simulations of the \ce{Ga2O3} system is now possible with the rapid advance of machine-learning (ML) interatomic potentials~\cite{deringer2021gaussian, friederich2021machine}.

Among many existing ML algorithms, one particularly well-established branch is the Gaussian approximation potential (GAP) based on the framework of Gaussian process regression~\cite{gap2010}.
The techniques of database construction, active/iterative training, and potential validation have been extensively studied and verified for single-element systems~\cite{bartok2018machine, rowe2020carbon, deringer2020a}, and oxides~\cite{sivaraman2020machine, erhard2022machine}.  
Currently, a ML-GAP~\cite{liu2020machine} and a deep neural-network potential~\cite{li2020a} were developed independently for perfect bulk $\beta$-\ce{Ga2O3} system with accurate prediction of its thermal properties.
A ML-GAP was recently developed for studying the thermal properties of the amorphous \ce{Ga2O3} system~\cite{liu2023unraveling}.
In our previous study, a ML-GAP was developed to describe two-dimensional phases of \ce{Ga2O3} with great accuracy, which successfully revealed specific kinetic pathways leading to phase transitions~\cite{SFzhao2021phase}. 
To date, however, there is few ML interatomic potential of extended generality that is required to describe the wide range of complex \ce{Ga2O3} bulk polymorphs and disordered structures.
Therefore, to meet the needs of the current experimental applications and extend the computational toolbox to larger spatiotemporal domains, we here develop two generalized ML-GAPs for simulations of all $\beta$, $\kappa$, $\alpha$, $\delta$, and $\gamma$ phases simultaneously. 
Moreover, we further extend their applicability to disordered, dispersed, and high-energy physical processes. 
This work opens up fundamental possibilities for computational exploration of large-scale dynamical processes in complex \ce{Ga2O3} systems.  

\section{Results}

\subsection{Generalized database and two GAP formalisms for \texorpdfstring{\ce{Ga2O3}}{} system} \label{sec:Database}

\begin{figure*}[ht!]
 \includegraphics[width=16cm]{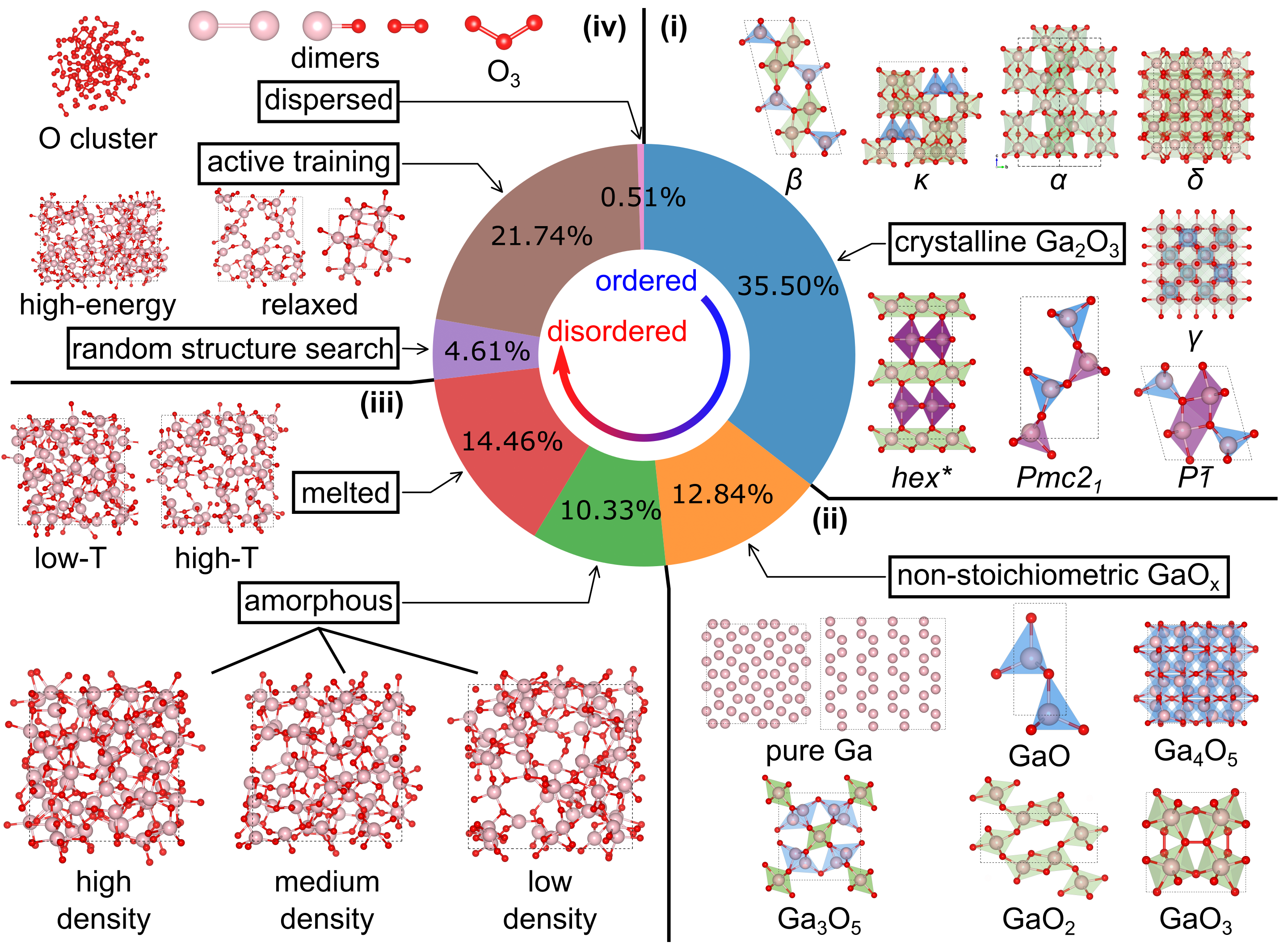}
 \caption{An overview of the DFT database.
 The fractions of the configuration-type-specific atom numbers to the total number (108,411) of atoms in the database are shown in the central part of the donut chart.
 For different configuration types, representative structures are shown as examples.
 In the panels (i) and (ii), the colored polyhedra within the crystalline structures show the 4-fold (blue)/5-fold (purple)/6-fold (green) \ce{Ga} sites.
 Note that the isolated Ga and O (not shown here) are also included in the database as the global references of potential energies.}
 \label{fig:database}
\end{figure*}

The fidelity of data-driven ML interatomic potentials heavily relies on the consistency and generality of the input data.
In our case, the consistency is guaranteed by well-converged accurate GGA-DFT calculations (see Sec.~\ref{sec:comput_method}, and Supplementary Appendix A), and the generality is ensured by a wide range of \ce{Ga}-\ce{O} structures as illustrated in Fig.~\ref{fig:database}, where a representative overview of the training database and its structures are shown. 
In total, 1,630 configurations with 108,411 atomic environments are included in the selected database.  
The potential energies, atomic forces, and virial stresses are stored for training of ML-GAPs.
As shown in Fig.~\ref{fig:database}, we arrange and classify the atomic configurations based on their degrees of physical (and/or chemical) ordering, starting from the perfect crystalline $\beta$-phase as the global energy minimum to the disordered high-energy structures. 

The four constituent parts of the database are shown in the respective panels of Fig.~\ref{fig:database}: (i) vital crystalline \ce{Ga2O3} polymorphs; (ii) non-stoichiometric \ce{GaO_{x}} including pure-\ce{Ga} phases; (iii) disordered bulk structures (amorphous) and melted (liquid) phases; and (iv) widely spread random structure search~\cite{deringer2020a}, structures from active training~\cite{sivaraman2020machine} and dispersed gaseous configurations.
Clearly, the part (i) with high-symmetry configurations is essential for modeling \ce{Ga2O3} polymorphs accurately, and the part (ii) is necessary to keep the overall chemical potential correct. 
Therefore, these two parts of the database are constructed and selected manually. 
In the part (i), we include the five experimentally identified polymorphs, and extend the data search to three computationally-predicted \ce{Ga2O3} phases, namely, the ``$Pmc2_{1}$" and ``$P\overline{1}$" phases from Ref.~\cite{wang2020discovery}, and the ``hex$^{*}$" phase from Ref.~\cite{swallow2021indium}.
In the part (ii), in addition to the experimentally known pure \ce{Ga} phases, the five computationally predicted \ce{GaO_{x}} lattices are selected from The Materials Project~\cite{jain2013commentary} database. 
On the other hand, the parts (iii) and (iv) consisting of low-symmetry, high-energy and dispersed configurations are stochastically generated to cover a wide range of atomic environments with the importance of diversity outweighing accuracy.
Another important aspect is that highly strained configurations are explicitly included in the database. 
For the parts (i) and (ii), all the bulk lattices are compressed and stretched uniformly with relaxed internal atomic positions for local minima under strain, and additional \textit{ab initio} MD (AIMD) to create randomized local atomic environments. 
For the disordered bulk structures in part (iii) and (iv), such as amorphous, melted and random structure search, the random cells are sampled with different densities, e.g., the densities of amorphous cells ranging from 3.8 to 5.4 g/cm$^{3}$ with seven uniformly increasing steps. 

Although the database is consistently calculated using GGA-DFT with the same level of \textit{ab initio} accuracy, even a diverse database is for any ML algorithm still prone to ``overfitting". 
To achieve high accuracy for all five \ce{Ga2O3} polymorphs and yet retain a smooth interpolation between the more disordered structures, one essential part of our ML training is a set of expected errors used for regularization ($\sigma$)~\cite{deringer2021gaussian}. 
These $\sigma$ values are chosen manually with systematical tests based on in-depth understanding of the physical nature of the atomic configurations. 
In general, small $\sigma$ values (0.002 eV for potential energy and 0.01 eV/\r A for force component) are set for vital crystalline polymorphs, while larger values ($0.0035\sim0.01$ eV, $0.035\sim0.1$ eV/\r A) are used for non-stoichiometric \ce{GaO_{x}} and disordered configurations.
Moreover, to enable the application of the developed potentials for ion irradiation simulations, we pay particular attention to the short-range interactions that take place in high-energy collision cascades (see Supplementary Appendix D).
We achieve the accurate description of interatomic repulsion by employing a set of external repulsive pair potentials explicitly fitted to the all-electron \textit{ab initio} data~\cite{nordlund1997repulsive} at short range. 
These repulsive potentials are included when training the GAP, so that only the difference between them and GGA-DFT has to be machine-learned. 
A well-behaved potential energy landscape from near-equilibrium to highly repulsive interatomic distances is ensured by also including high-energy structures in the GGA-DFT training database~\cite{byggmastar2019machine}.
More detailed information of the database construction and the GAP training process is provided in Supplementary Appendices B, C, and D. 

One major advance in our ML-GAPs is the choices of local-atomic-environment descriptors. 
In the GAP framework, the contributions to the total machine-learned energy prediction can be supplied by multiple terms using different descriptors~\cite{deringer2021gaussian}.
The commonly-used combination of a two-body (2b) and the high-dimensional smooth overlap of atomic positions (SOAP) descriptor can distinguish different atomic configurations with high sensitivity and leads to excellent accuracy~\cite{soap2013, twobody2020}. 
However, the computational cost of soapGAP is fairly high, which limits the accessible spatiotemporal domain of MD simulations (with reasonable computational effort) only to the order of 10$^{4}$ atoms and 1 ns. 
To extend the length and time scales of an MD simulation to millions of atoms and tens of nanoseconds, we also employ the recently developed framework of tabulated low-dimensional GAPs (tabGAP)~\cite{byggmastar2022simple}. 
In tabGAP, a combination of only low-dimensional descriptors is used (2b+3b+EAM). Here, 3b refers to the three-body cluster descriptor~\cite{bartok2015gaussian} and EAM is a descriptor corresponding to the pairwise-contributed density used in embedded atom method potentials~\cite{byggmastar2022simple} (see Supplementary Appendix C for details). 
The low dimensionality of the descriptors ultimately limits the flexibility, and hence also the accuracy, of the interatomic potential. 
However, the advantage is that the low dimensionality allows for tabulating the machine-learned energy predictions onto one- and three-dimensional grids, so that the energy can be computed with efficient cubic-spline interpolations rather than the Gaussian process regression of GAP. 
A speedup of two orders of magnitude is achieved by this tabulation.

With these strategies, we propose two versions of ML-GAPs: (1) a soapGAP that is very accurate but computationally slow ($\sim8\times10^{2}$ MD steps/(s$\cdot$atom)) and (2) a tabGAP that has lower accuracy compared to the soapGAP, but much higher computational efficiency ($\sim3.2\times10^{5}$ MD steps/(s$\cdot$atom)). 
The former is suitable for detailed medium-scale MD simulations and the latter, which is 400 times faster than soapGAP (more details in Supplementary Appendix G) is advisable for large-scale or long-time MD simulations, including high energy collision cascade simulations.

\begin{figure*}[ht!]
 \includegraphics[width=16cm]{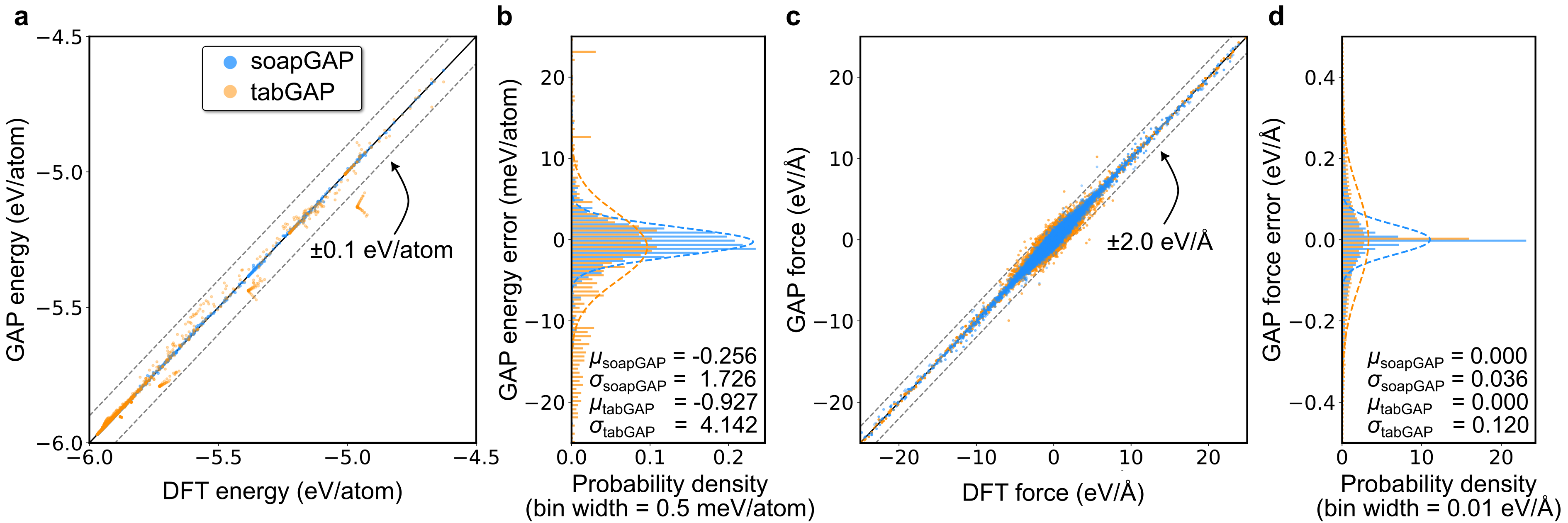}
 \caption{The validation of the soapGAP and tabGAP: 
 Scatter plots of (a) energies and (c) force components versus DFT data, and the corresponding histogram plots of the probability densities of the errors. 
 Note that the validated range of the energies and forces in (a) and (c) are two orders of magnitude larger than the span of the errors in (b) and (d).
 The distributions of the errors are fitted with the Gaussian distribution with the means ($\mu$) and standard deviations ($\sigma$) listed in (b) and (d).  
 This validation is done for all the important solid-bulk configurations, i.e., crystalline \ce{Ga2O3}, non-stoichiometric \ce{GaO_{x}}, amorphous, and random structure search, with 1,202 energies and 205,830 force components in total.
 }
 \label{fig:accuracy}
\end{figure*}

The validation of the soapGAP and tabGAP is shown in Fig.~\ref{fig:accuracy}, where both potential energies and force components are plotted against the reference DFT data. 
The potential energies predicted by soapGAP follow tightly the DFT calculations in the entire range from the global minimum ($E_\mathrm{p}$($\beta$-\ce{Ga2O3})=$-$5.969 eV/atom) to the $-$4.5 eV/atom (well above the average energies of the high-temperature liquid phases).
All points lie along the solid diagonal line in Fig.~\ref{fig:accuracy}a within the two dashed lines with the reference deviation of $\pm$0.1 eV/atom, indicating an excellent agreement with the DFT data. 
The tabGAP-predicted points follow the the diagonal line tightly from $-$5.969 to $-$5.750 eV/atom, but have several grouped points deviating from the line in the higher energy range.
Indeed, the soapGAP yields better accuracy and precision comparing to the tabGAP as further confirmed by the standard deviation of the energy error distribution in Fig~\ref{fig:accuracy}b, where $\sigma_\mathrm{tabGAP}\simeq2.4\sigma_\mathrm{soapGAP}$. 
This is expected, because the high-dimensional SOAP descriptors can better distinguish the small differences in atomic environments, especially for element-specific differences.  
Nevertheless, the tabGAP shows good accuracy for the important crystalline configurations.
We note that one particular difficulty for tabGAP is the non-stoichiometric \ce{GaO} phases which are shown as the grouped points outlying the $\pm$0.1 eV/atom reference zone around $-$5.0 eV/atom in Fig~\ref{fig:accuracy}a. 
However, this issue is not critical for the general utility of the tabGAP, as demonstrated later.   
More detailed notes of the GAP-predicted non-stoichiometric \ce{GaO_{x}} phases is included in Supplementary Appendix E.   
Second, the validation of force components indicates a comparable trend between the soapGAP and tabGAP as shown in Figs.~\ref{fig:accuracy}c and d. 
The tabGAP-predicted points are slightly more scattered than the soapGAP-predicted ones, with a very small number of outlying points far from the $\pm$2.0 eV/\r A reference zone. 
The standard deviations of the force error distribution are $\sigma_\mathrm{tabGAP}\simeq3.33\sigma_\mathrm{soapGAP}$, similar to that of the energy errors. 

We note that both the soapGAP and tabGAP errors are two orders of magnitude smaller than the validation span for both potential energies and force components. 
These overall balanced small errors, and more importantly, correct physical predictions are optimized and achieved with many iterations of benchmark training and testing. 
For the two ML-GAPs, our primary aim is to achieve good accuracy for the \ce{Ga2O3} polymorphs, while retaining physically reasonable predictions in a wide region of the configuration space. 
The raw energy and force component errors tell little about the actual physically important properties. 
For example, the energy difference between some of the polymorphs is very small, so that even with a small error in the energy one might still have an incorrect order of stability for the crystal phases, such as a false global minimum different from the $\beta$-phase.
Therefore, we continue testing the soapGAP and tabGAP in more realistic physical scenarios in the next sections.   

\subsection{Five individual polymorphs} \label{sec:Polymorphs}

We begin the demonstration of generalized accuracy of our interatomic potentials by discussing the GAP-predicted structural properties of the five \ce{Ga2O3} polymorphs. 
The ground-state lattice parameters of the five polymorphs are listed in Table~\ref{tab:lattice}. 
Strikingly, for these multi-symmetry and diverse configurations, both the soapGAP and tabGAP are in a great agreement with DFT, with the largest error being only 0.35\% from DFT values (corresponding to the $\kappa$-phase, tabGAP-predicted $a$ values).

\begin{table}[ht!] 
\caption{Lattice parameters (\r A) and angle $\beta$ ($\degree$) for five polymorphs predicted by GGA-DFT, soapGAP and tabGAP. 
The experimental (Exp.) data are taken from Refs.~\cite{ahman1996a,SFzhang2021temperature} (for $\beta$), \cite{cora2017real} (for $\kappa$), \cite{marezio1967bond} (for $\alpha$), and \cite{kato2023demonstration} (for $\delta$), and \cite{playford2014characterization} (for $\gamma$).
The $\gamma$-phase is a unique defective spinel-type structure with inherent disorder~\cite{ratcliff2022tackling, mu2022phase}, so the exact lattice parameter is not well-defined, and an averaged value over three random cells is shown here. 
The parameters for the $\beta$- and $\alpha$-phases are for the 20- and 30-atom conventional cells, respectively.  
}
\label{tab:lattice}
\begin{ruledtabular}
\begin{tabular}{l r r r r}
                   & Exp.     & DFT              & soapGAP     & tabGAP      \\
\colrule
\multicolumn{5}{l}{$\beta$-phase (Monoclinic, $C2/m$)}                               \\
$a$                & 12.214   & 12.460           & 12.480      & 12.482      \\
$b$                & 3.037    & 3.086            & 3.086       & 3.084       \\
$c$                & 5.798    & 5.879            & 5.871       & 5.885       \\
$\angle \beta$     & 103.832  & 103.680          & 103.600     & 103.880     \\
\hline
\multicolumn{5}{l}{$\kappa$-phase (Orthorhombic, $Pna2_{1}$)}                        \\
$a$                & 5.046    & 5.126            & 5.141       & 5.144       \\
$b$                & 8.702    & 8.802            & 8.780       & 8.792       \\
$c$                & 9.283    & 9.419            & 9.413       & 9.419       \\
\hline
\multicolumn{5}{l}{$\alpha$-phase (Hexagonal, $R\overline{3}c$)}                      \\
$a=b$              & 4.982    &   5.064          & 5.066       & 5.067       \\
$c$                & 13.433    &   13.632         & 13.634      & 13.610      \\
\hline
\multicolumn{5}{l}{$\delta$-phase (Cubic, $Ia\overline{3}$)}                         \\ 
$a=b=c$            & 9.255    & 9.411            & 9.411       & 9.415       \\
\hline
\multicolumn{5}{l}{$\gamma$-phase  (Cubic*, $Fd\overline{3}m$)}                      \\
$\overline{a} \simeq \overline{b} \simeq \overline{c}$  
                                    & 8.238 & $\sim8.370$ & $\sim8.360$ & $\sim8.355$      \\
\end{tabular}
\end{ruledtabular}
\end{table}

\begin{figure*}[ht!]
 \includegraphics[width=16cm]{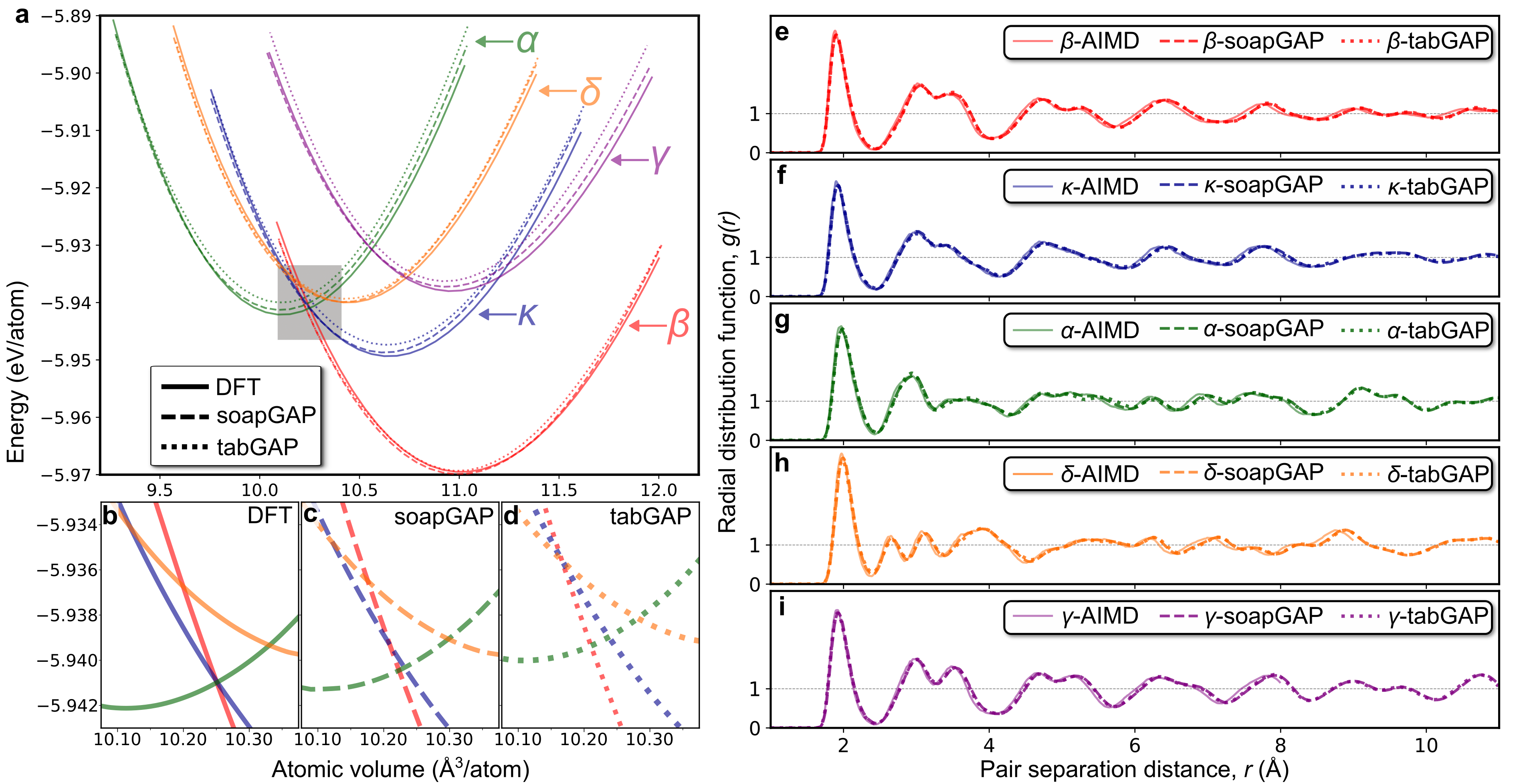}
 \caption{
 Left panel: Comparison of the DFT, soapGAP and tabGAP energies for the five experimentally identified $\beta/\kappa/\alpha/\delta/\gamma$ polymorphs.
 In the overall plot (a), the gray shadow region corresponds to the energy-crossing region magnified and compared closely in (b,c,d).
 The energy difference between DFT and soapGAP data is less than 2 meV/atom.
 Right panel: RDFs, $g(r)$, of $NPT$-MD simulations at 900 K and 0 bar for the five \ce{Ga2O3} polymorphs (e-i), respectively. 
 The DFT AIMD simulations are run with small $120\sim160$-atom cells, hence the cutoff distances of the AIMD $g(r)$ are set based on the shortest side length of the corresponding cells.
 }
 \label{fig:bulk}
\end{figure*}

This great agreement of the structural properties also extends to strained systems, as shown in Fig.~\ref{fig:bulk}a. 
The energies are plotted against the atomic volumes under isotropic strain with the internal atomic positions relaxed to the corresponding local minima. 
The soapGAP- and tabGAP-predicted energies follow the harmonic curves nicely with a small energy drift of $2\sim3$ meV from the DFT curves, spanning a wide range of volumetric strain from 91.27\% to 109.27\% (lattice strain from $-3$\% to 3\%). 
This also means that the GAP-predicted bulk moduli of these polymorphs follow the similar trend of DFT data with acceptable deviation (see Supplementary Appendix F).
The order of phase stability is $\beta < \kappa < \alpha < \delta < \gamma$, the same as the recent accurate hybrid-functional DFT data in Ref.~\citenum{mu2022phase}, and the GGA-DFT data in this work as well. 
We now focus on an intriguing $\beta$/$\kappa$/$\alpha$/$\delta$ energy-degenerate ($\pm$8 meV/atom) region (see the shadowed box in Fig.~\ref{fig:bulk}a), which attracted much attention in a number of recent studies~\cite{anber2020structural, cora2020insitu, azarov2022disorder, lion2022elastic, azarov2023universal}. 
This region is reported to be a thermodynamically balanced point of $\alpha\rightarrow\beta$/$\kappa\rightarrow\beta$ solid-state phase transitions occurring when the \ce{Ga2O3} system is under pressure of $24\sim44$ GPa. 
Here, we note that the energy-volume curve of the $\delta$-phase lies close to the $\beta$/$\kappa$/$\alpha$ curves in this region, so it is included in the comparison. 
As shown in Figs.~\ref{fig:bulk}b-d, the $\beta$/$\kappa$/$\alpha$ balanced point is accurately reproduced by soapGAP with a marginal relative shift of 2 meV/atom in energy.
The fine relative energy-volume balance is almost identical to the DFT curves.  
On the other hand, the tabGAP-predicted energy drifts are larger than the ones from the soapGAP, however, the overall energy-degenerate region lies well within a small energy-volume range of $\Delta E=\pm$4 meV/atom and $\Delta V= \pm$0.05 \r A$^{3}$/atom. 
Given the fact that the average atomic kinetic energy at 300 K is $\sim38.8$ meV/atom, the tabGAP-predicted $\Delta E/\Delta V$ is a good approximation of the energy-degenerate point.  

One essential aim of using our interatomic potentials is to attain \textit{ab initio} accuracy with temporal and spatial limits which cannot be accessed with DFT.
Therefore, we further show the radial distribution functions (RDFs), $g(r)$, in Fig.~\ref{fig:bulk} for the five polymorphs from isobaric-isothermic ($NPT$) MD at 900 K and 0 bar.
The soapGAP and tabGAP MD simulations are run with $1000\sim2000$-atom cells, while the AIMD simulations are run with $120\sim160$-atom cells.    
As shown in Figs.~\ref{fig:bulk}e-i, both soapGAP and tabGAP MD runs lead to almost indistinguishably similar RDFs as compared to the AIMD runs, with a marginal scaling mismatch in the pair distance longer than 4 \r A. 
Similarly for all the five polymorphs, the first peaks of the RDFs at $\sim2$ \r A correspond to the average \ce{Ga}-\ce{O} bond length, whereas the long-range peaks vary depending on the symmetry of oxygen stacking and Ga-site occupation arrangement. 

\begin{figure*}[ht!]
 \includegraphics[width=16cm]{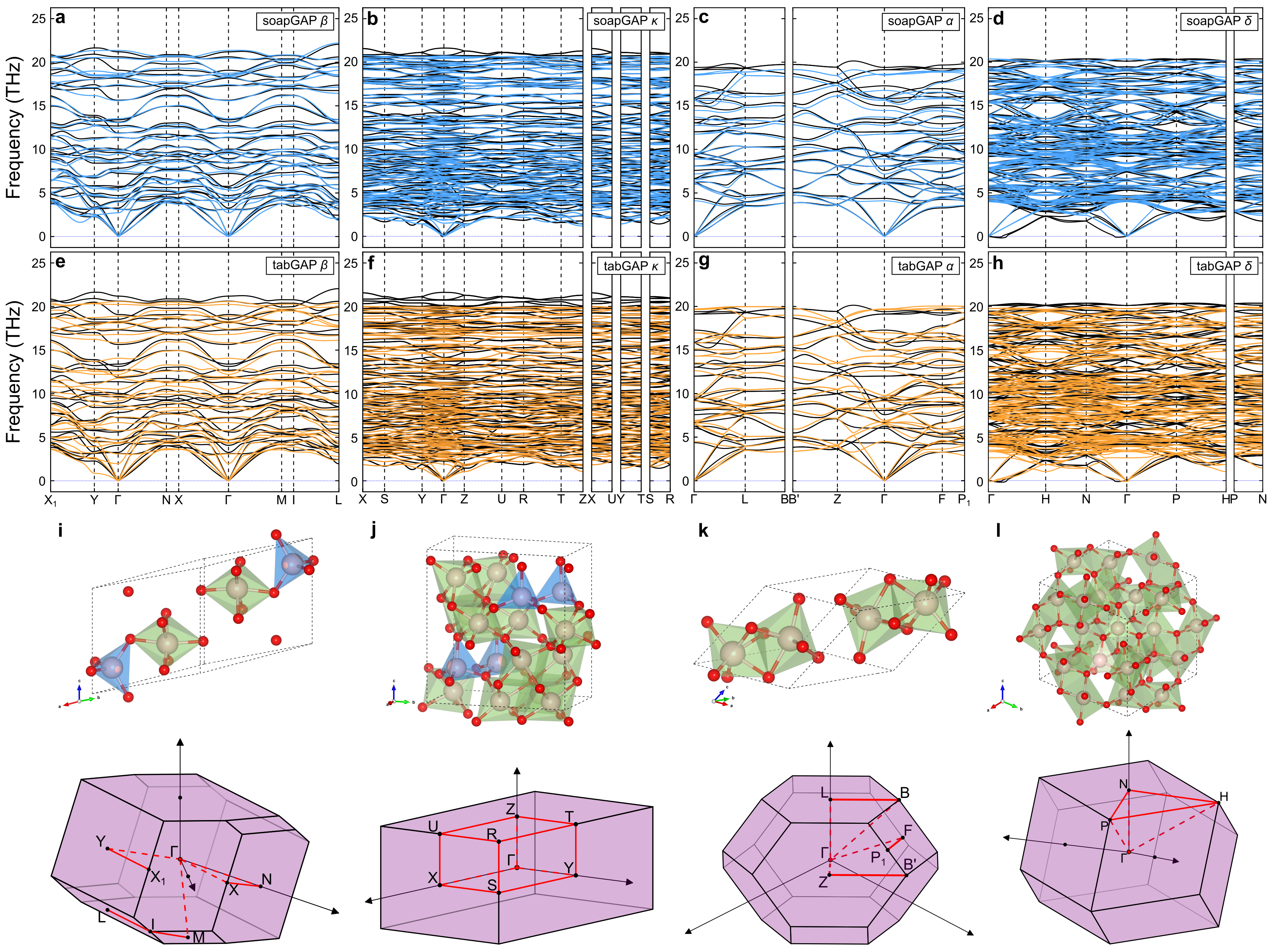}
 \caption{
 (a-h) Phonon dispersion curves for the $\beta/\kappa/\alpha/\delta$-\ce{Ga2O3} phases compared with GGA-PBE-DFT calculations. 
 The black, blue, and orange bands are from DFT, soapGAP, and tabGAP, respectively. 
 The non-analytical-term correction of the longitudinal optical (LO)-transverse optical (TO) splitting in an ionic solid at the $\Gamma$ point can be further included with the DFT-calculated Born effective charge which are currently not included here for a consistent comparison. 
 (i-l) The corresponding $\beta/\kappa/\alpha/\delta$ unit cells (upper) and the first Brillouin zones of the reciprocal lattices labelled with the high-symmetry k-points (lower).
 }
 \label{fig:phonon}
\end{figure*}

To assess the ability of our potentials to predict thermal properties of the different polymorphys of \ce{Ga2O3}, we calculate the phonon dispersion curves of the $\beta/\kappa/\alpha/\delta$ polymorphs using both potentials and compare these curves against the results obtained directly with the DFT method. 
The results are shown in Fig.~\ref{fig:phonon}. 
Compared to the $\beta$-\ce{Ga2O3} phonon dispersion calculated using the specific $\beta$-\ce{Ga2O3} ML-GAP potential~\cite{liu2020machine}, we see that our general-purpose soapGAP is capable to reproduce most of the analytical phonon branches with only slightly larger deviation (Fig.~\ref{fig:phonon}a). 
As expected, the tabGAP-predicted phonon dispersion curves deviate more from the DFT curves, however, all results follow closely the expected trends and are well within the expected whole phonon band, with one exceptional deviation in the lowest acoustic branch at the $Y$ point (Fig.~\ref{fig:phonon}e). 
Overall, a similar trend is seen for the other polymorphs. 
Given the fact that the phonon dispersion and thermal properties of the \ce{Ga2O3} polymorphs exhibit high complexity and anisotropy, more specifically trained potentials with specialized database are the suitable choice to improve the accuracy for this purpose.

\begin{figure*}[ht!]
 \includegraphics[width=16cm]{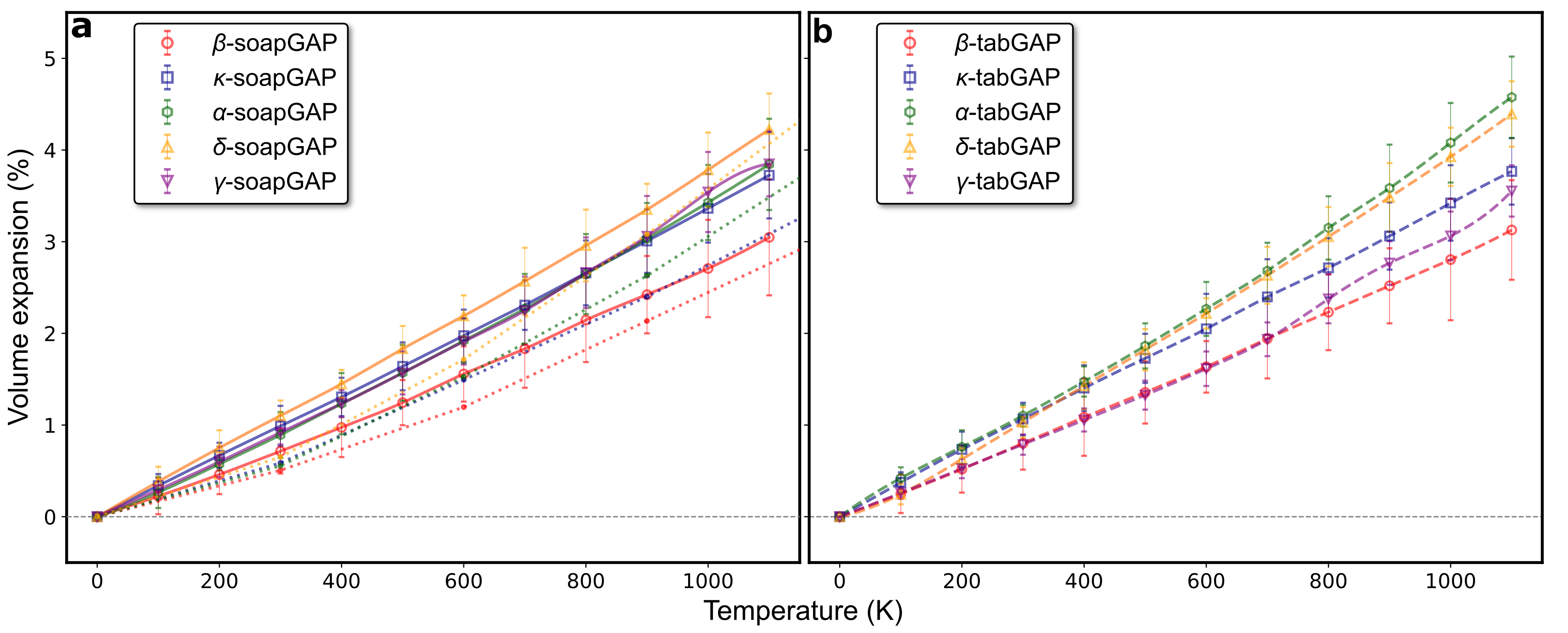}
 \caption{
 The order of the thermal expansion for the four polymorphs predicted by (a) soapGAP: $\beta < \alpha \simeq \kappa \simeq \gamma < \delta$; and (b) tabGAP: $\beta \simeq \gamma < \kappa < \delta < \alpha$. 
 The order predicted by the soapGAP agrees well with the first-principles calculations in Ref.~\citenum{yoshioka2007structures} (the four dotted lines for $\beta/\kappa/\alpha/\delta$ in panel (a) with the same colors as the soapGAP ones). 
 The main differences between the soapGAP and tabGAP are the $\alpha$-phase and $\gamma$-phase.
 The error bars represent the standard deviations caused by temperature/pressure fluctuation.  
 }
 \label{fig:thermal}
\end{figure*}

As the last test of properties of specific polymorphs, we run $NPT$ MD simulations at zero pressure and different temperatures to obtain the thermal expansion curves.
As shown in Fig.~\ref{fig:thermal}, both the soapGAP and the tabGAP predict the lowest expansion rate for the $\beta$-phase, but a different order for the remaining four polymorphs. 
The absolute values of the $\beta$/$\kappa$/$\delta$-phases are very similar between the soapGAP and tabGAP.  
The main differences occur in the $\alpha$- and $\gamma$-phases where the tabGAP slightly overestimates the thermal expansion of $\alpha$-phase, and underestimates the $\gamma$-phase. 
Moreover, the $\gamma$-\ce{Ga2O3} at elevated temperatures should be treated with a special caution as both of the thermal-expansion curves for this phase deviate from the parabolic function form under such condition.
This is likely because the $\gamma$-phase is a disordered defective spinel lattice. 
The average atomic volume is close to the $\beta$-phase (Fig.~\ref{fig:bulk}) and the sub-lattice of oxygen follows face-centered-cubic (fcc) stacking in both the $\beta$- and $\gamma$-phases, which may explain why the less flexible tabGAP cannot distinguish the thermal expansion of these two phases.
The experimental reference of thermal expansion of the \ce{Ga2O3} polymorphs is not very extensive, however, the first-principles calculation~\cite{yoshioka2007structures} is in the same order as shown here for soapGAP. 

\subsection{Disordered liquid and amorphous structures} \label{sec:Disordered}

As general-purpose interatomic potentials, the capabilities of reproducing the disordered structures with \textit{ab initio} accuracy are vital for complex high-temperature applications. 
Liquid and amorphous phases of the \ce{Ga2O3} are of tremendous interests for experimental studies and applications, such as edge-defined film-fed growth and thermally-driven crystallization. 

It is well-known that the amorphous phase is physically defined as a material state with present short-range ordering and absent long-range ordering.
Specifically for the amorphous \ce{Ga2O3} system, the short-range ordering is predominated by the highly ionic nature of the Ga-O bond (the Pauling’s ionicity of \ce{Ga2O3}, $\sim0.49$~\cite{ma2016intrinsic}). 
Therefore, the short-range ordering can be considered as the high-symmetry localized tetrahedral (4-fold) and octahedral (6-fold) \ce{Ga} sites with possible over- and under-coordinated sites.
On the other hand, the long-range ordering of a given \ce{Ga2O3} system relies on the synergetic ordering of the \ce{Ga} and \ce{O} sub-lattices.
The \ce{O} sub-lattices in crystalline polymorphs follow close-packed stacking orders, e.g., $\beta$-\ce{O} in fcc, $\kappa$-\ce{O} in $4H$-ABCB, and $\alpha$-\ce{O} in hcp. 
The \ce{Ga} sub-lattice further determines the ratio of the 4-fold to the 6-fold sites in different polymorphs~\cite{swallow2020influence}.
Therefore, depending on the ordering of the \ce{Ga}/\ce{O} sub-lattices, a \ce{Ga2O3} system can be classified into three types: (i) perfect polymorphs; (ii) defective cell; and (iii) disordered cell. 
The types (i) and (iii) are easily understood with conventional definitions, while the type (ii) is specifically defined here as a \ce{Ga2O3} system that has long-range ordered \ce{O} sub-lattice in close-packed stacking and rather random \ce{Ga} sub-lattice with low symmetry. 
This aspect will be discussed in detail in Sec.~\ref{sec:Transition}.

\begin{figure*}[ht!]
 \includegraphics[width=16cm]{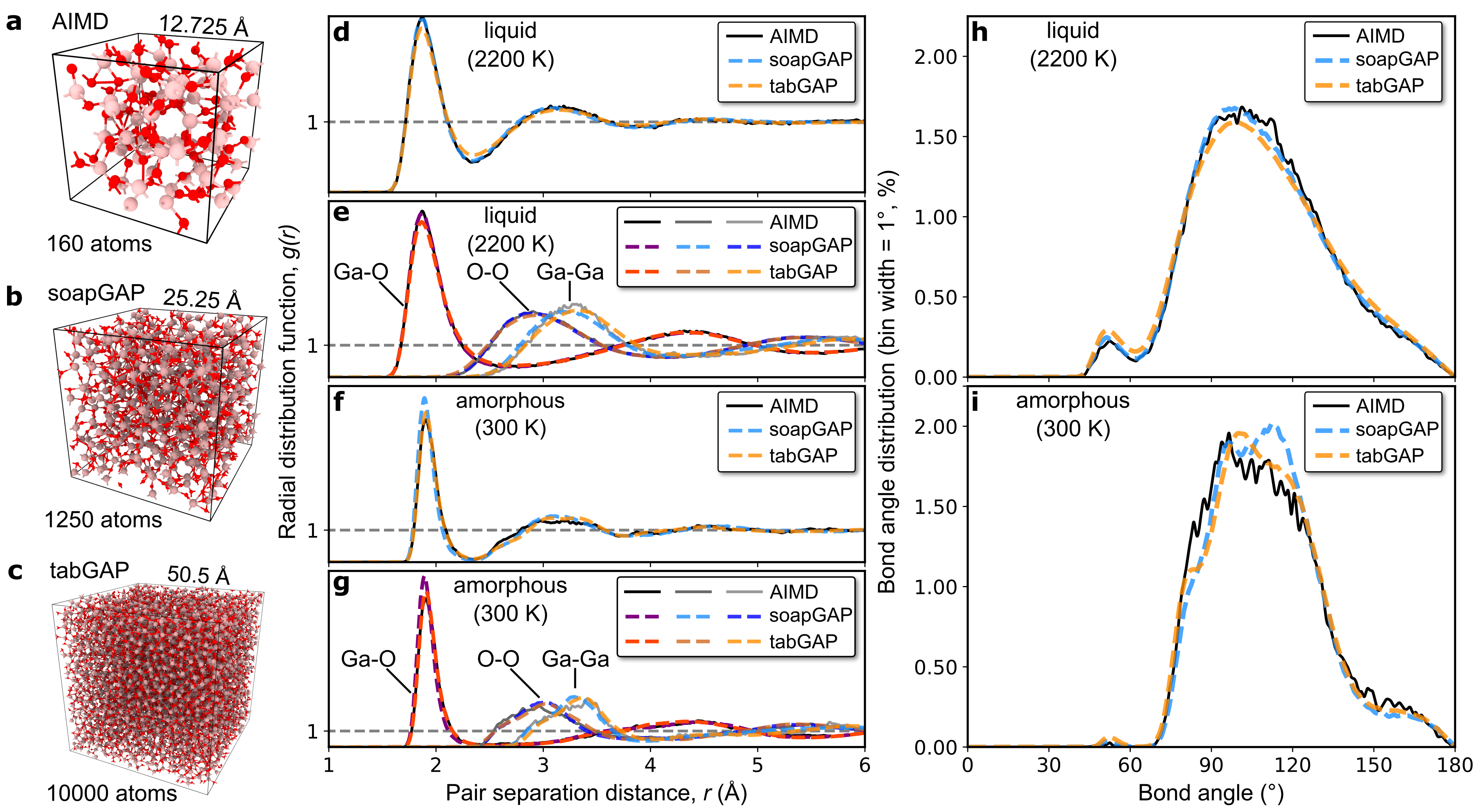}
 \caption{Left panel: Three liquid/amorphous cells with consistent density of 4.84 g/cm$^{3}$ used for (a) AIMD, (b) soapGAP and (c) tabGAP runs. 
 Middle panel: RDFs and PRDFs of the (d,e) liquid and (f,g) amorphous configurations. 
 Right panel: Bond angle distributions of the \ce{Ga}-\ce{O} bonds with the cutoff distance of 2.4 \r A where the first valley of the RDF lies.}
 \label{fig:gr_melted}
\end{figure*}

In the following, we analyze the performance of the soapGAP and tabGAP for homogeneous disordered liquid-amorphous \ce{Ga2O3} structures. 
We perform these MD simulations in canonical ($NVT$) ensemble at 2200 K using a 10000-atom 50.5-\r A$^{3}$ cubic cell for tabGAP, a 1250-atom 25.25-\r A$^{3}$ for soapGAP, and a 160-atom 12.725-\r A$^{3}$ cubic cell for AIMD. 
The densities of all three cells are the same 4.84 g/cm$^{3}$, matching the experimentally reported density of liquid \ce{Ga2O3} at 2123 K~\cite{dingwell1992density}. 
The amorphous \ce{Ga2O3} system was simulated in a wide-spread range of densities $3.9\sim5.3$ g/cm$^{3}$~\cite{yu2003growth}. 
Exemplary snapshots of the three cells are illustrated in Figs.~\ref{fig:gr_melted}a-c.

As shown in Figs.~\ref{fig:gr_melted}d-i, we validate the soapGAP and tabGAP for both liquid and amorphous structures by analyzing the RDFs, partial RDFs (PRDFs) and bond-angle distributions. 
Strikingly, both the soapGAP and tabGAP describe the liquid structures in excellent agreement with the AIMD reference.
For the RDFs, the short-range fingerprint of the liquid structure is the ratio of the first peak to the second peak. 
Moreover, the element-pair-wise PRDFs can reveal whether these peaks are partially contributed by more than one types of bonds.
As clearly seen in Fig.~\ref{fig:gr_melted}e, the first peaks of the liquid RDFs (Fig.~\ref{fig:gr_melted}d) are fully composed of \ce{Ga}-\ce{O} bonds, indicating no chemical segregation in the liquid phase. 
The first peak of the tabGAP RDF is slightly smaller than those of the AIMD and soapGAP RDFs, indicating slightly lower coordination of \ce{Ga} atoms in the tabGAP liquid system. 
The second peak is composed of the merged peaks of the first-nearest \ce{O}-\ce{O} and \ce{Ga}-\ce{Ga} neighbors, which is again in a very good agreement between the results of all three cells. 
Moreover, we also see in Fig.~\ref{fig:gr_melted}h that the \ce{Ga}-\ce{O} bond-angle distributions agree very closely for all three methods. 
The presented comparisons indicate that the short-range \ce{Ga}-\ce{O} interactions within the first RDF peak are captured accurately by both soapGAP and tabGAP.  

The amorphous $NVT$ MD simulations are conducted at 300 K and 1 bar pressure by quenching the corresponding liquid systems. 
The quenching rates are set to 170, 34, and 3.4 K/ps, for the AIMD, soapGAP, and tabGAP runs, respectively, and the analyses are conducted with additional 300-K $NVT$ simulations after the quenching process.
As shown in Fig.~\ref{fig:gr_melted}f and g, the first peaks of the RDFs of amorphous structures are stronger and narrower than those of the liquid phase, with nearly-zero valleys between the first and second peaks (note the different scales between Fig.~\ref{fig:gr_melted}d and f).  
Very shallow third peaks appear around 4.5 \r A corresponding to the second shells of \ce{Ga}-\ce{O} PRDFs. 
The overall agreement of all three methods is excellent again.
Only minor differences are observed at the height of the first peaks of RDFs in soapGAP and tabGAP, and shifted maxima of the bond-angle distributions (Fig.~\ref{fig:gr_melted}i). 
These two differences indicate a small deviation in the short-range \ce{Ga}-\ce{O} bonding configurations. 
However, some key trends in the bond-angle distributions, such as the vanished signals around $60\sim70\degree$ and the lower shoulder at 150$\degree$, are predicted to be exactly the same as the AIMD results. 
The comparison of fine details in the distributions should be made with caution, since there is a fifty times difference in quenching rate for the three amorphous runs. 
It is highly probable that some slow atomic movements are not caught in AIMD, which is run on a $\sim10$-ps time scale. 
We note that our soapGAP-predicted bond-angle distribution is similar to the results obtained with the soapGAP-relaxed cell of the same density reported in Ref.~\cite{liu2023unraveling}. 
Furthermore, a test result shows our potentials can lead to a similar topological ring-size statistics which were reported in Ref.~\cite{liu2023unraveling} as well, which is worthy for a future systematical investigation.
Therefore, we conclude that both the soapGAP and tabGAP can reproduce the important features of the disordered \ce{Ga2O3} system with high accuracy.
This fact encourages us to further employ the interatomic potentials to the complex heterogeneous liquid-solid interface system.

\subsection{Liquid-solid phase transition} \label{sec:Transition}

We now utilize the developed interatomic potential to examine the liquid-solid ($\beta$) phase transition process which is crucial for the synthesis of \ce{Ga2O3} via liquid-phase growth methods~\cite{galazka2022two, heinselman2022projected} and must be understood fundamentally at the atomic level. 

\begin{figure*}[ht!] 
 \includegraphics[width=16cm]{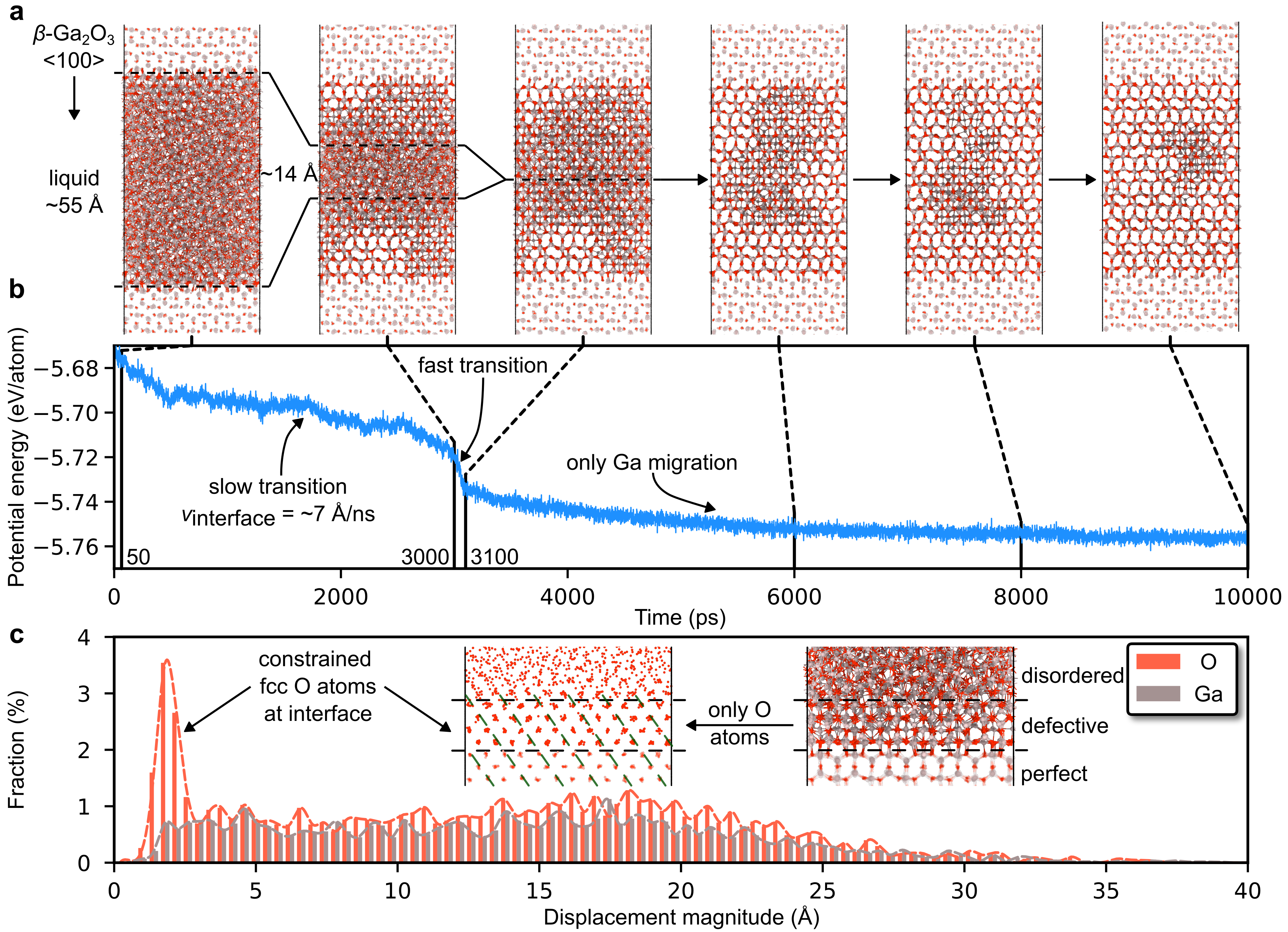}
 \caption{Liquid-solid phase transition simulation running with the tabGAP at 1500 K. 
 (a) Evolution of atomic configurations viewed from [010] direction. 
 The \ce{O} and \ce{Ga} are in red and brown, respectively. 
 (b) Potential energy: three distinct regimes are identified. 
 (c) Analysis of the atomic displace magnitude (between the 50-ps and 3000-ps frames). 
 A group of unusual low-mobility \ce{O} atoms are revealed. 
 This group is corresponding to the rapid formation of the fcc \ce{O} atoms at the defective region of the interface, as illustrated in the inset atomic configurations.}
 \label{fig:transition}
\end{figure*}

For the tabGAP simulation, we join vertically a perfect 5760-atom $\beta$-phase slab with a (100) top surface and a pre-thermalized disordered 5760-atom \ce{Ga2O3} cell. 
The periodic boundary conditions are applied in all directions, as illustrated in Fig~\ref{fig:transition}a. 
This way we simulate two liquid-solid interfaces simultaneously in one simulation cell of the size $\sim36\times36\times100$ \r A$^{3}$.
The two interfacial regions are first relaxed to a local energy minimum to avoid atoms overlapping.
We then run $NPT$ MD at 1500 K and 1 bar for 10 ns. 
The evolution of the atomic structure correlated with the change of potential energy per atom is shown in Fig. \ref{fig:transition}(a-b). 
Intriguingly, we can distinguish three different stages of the phase transition process that are labeled in Fig.~\ref{fig:transition}b as ``slow transition", ``fast transition" and ``only Ga migration".
The initial slow transition is characterized by heterogeneous nucleation of the ordered phase within the liquid near the interface which is followed by the spontaneous self-assembling of \ce{O} into a fcc stacking, while \ce{Ga} tends to occupy randomly the tetra- and octahedral sites. 
This slow transition propagates with the rate of $\sim7$ \r A/ns from the liquid-solid interface.  

After 3000 ps, the thickness of the liquid layer is down to $\sim14$ \r A, which is the critical point when the fast transition phase begins. 
A distinct first-order transition finishes within 100 ps, and results in a completely ordered \ce{O} sub-lattice and defective \ce{Ga} arrangement.  
Then, the subsequent process only involves migration of \ce{Ga} atoms, as can be seen in the snapshots at 6000/8000/10000 ps in Fig.~\ref{fig:transition}a, where the defective \ce{Ga} atoms gradually recover to perfect $\beta$-phase and have a collective mobility as a defect cluster.         

For quantitative analysis of the slow transition process, we plot the distribution of atomic displacements comparing the simulation cell snapshots at 50 ps and 3000 ps. 
The histogram includes all the initially liquid atoms, as shown in Fig.~\ref{fig:transition}c.
The overall displacement distributions are fairly uniform and similar for \ce{O} and \ce{Ga} atoms at the large magnitudes ($5\sim40$ \r A), suggesting that a large fraction of \ce{Ga} and \ce{O} atoms have similar mobility.
However, a peculiar peak appears at 2 \r A in the distribution of \ce{O}, indicating a highly constrained group of \ce{O} atoms. 
We identify this group of \ce{O} atoms by closely tracking all the \ce{O} atom trajectories, and find that it corresponds exactly to the initial formation of the defective layer at the interface.
As shown in the inset figure in Fig.~\ref{fig:transition}c, the \ce{O} atoms quickly align to fcc stacking (dashed green lines guide the eye) and stay constrained locally throughout the whole transition process. 
Rather counter-intuitively, this solid-confined low mobility does not occur for the \ce{Ga} atoms, as the displacement distribution of \ce{Ga} spread evenly in a wide range of values, indicating that the mobility of \ce{Ga} atoms is less constrained by the solidification process. 
In the view of the entire process, this is expected, as the further recovery of the defective layer proceeds only via the movements of mobile \ce{Ga} atoms.

As a final remark, we note that although the phase transition simulations were performed with tabGAP, the same promotion of fcc stacking alignment and local confinement of the interfacial \ce{O} atoms is seen in the soapGAP simulations as well. 
In principle, this curious phenomenon  could be verified and studied experimentally with \textit{in-situ} scanning transmission electron microscopy or other atomic-level imaging techniques.     

\section{Conclusion} \label{sec:Con}

\ce{Ga2O3} is emerging as a promising semiconductor material for industrial applications, but its structural complexity with many stable polymorphs makes it an extremely challenging material to model in large-scale atomistic simulations.
We have developed two versions of generalized ML-GAP interatomic potentials for the \ce{Ga2O3} system, soapGAP and tabGAP, offering different balances between computational speed and accuracy.
Our results demonstrate that both interatomic potentials are capable of describing the five $\beta$/$\kappa$/$\alpha$/$\delta$/$\gamma$ stable polymorphs as well as the amorphous and liquid phases with high accuracy.
The simulation of the liquid-solid phase transition reveals the fast formation of the constrained fcc \ce{O} atoms at the interfacial defective layer followed by the slow migration of \ce{Ga} atoms.
The \ce{Ga2O3} database of structures developed in this work can be readily transferred and used as input data for other ML frameworks.
In a broad perspective, our interatomic potentials together with the highly intensive ongoing experimental investigations will enable atom-level design of new \ce{Ga2O3}-based applications.

\section{Computational Methods} \label{sec:comput_method}

The DFT calculations were conducted using the Vienna \textit{Ab initio} Simulation Package (VASP)~\cite{vasp1993}, employing the projected augmented-wave (PAW) method~\cite{paw1994} with 13 ($3d^{10}4s^{2}4p^{1}$) and 6 ($2s^{2}2p^{4}$) valence electrons for Ga and O, respectively. 
The Perdew-Burke-Ernzerhof (PBE) version~\cite{pbe1996} of the generalized gradient approximation was used as the exchange-correlation functional.
In the DFT calculations, the electronic states were expended in plane-wave basis sets with an energy cutoff of 700 eV.
The Brillouin zone was sampled with $\Gamma$-centered k-mesh grids with a maximum spacing of 0.15 \r A$^{-1}$ which was equivalent to a dense 3 $\times$ 12 $\times$ 6 grid for a monoclinic 12.461 $\times$ 3.086 $\times$ 5.879 \r A unit cell. 
Gaussian smearing with a width of 0.03 eV was used to describe the partial occupancies of the electronic states. 
The detailed convergence tests on the plane-wave energy cutoff, and the k-mesh grid are attached in Supplementary Appendix A.
We chose 10$^{-6}$ eV and 5 $\times$ 10$^{-3}$ eV/\r A as the energy and force convergence criteria for the optimization of the electronic and ionic structures, respectively. 
We note that the high accuracy and consistency of the electronic energy and force sampling is necessary to guarantee the energy consistency of DFT database. 
This consistency is essential for constructing smooth potential surfaces with GAP and both 2b+SOAP and 2b+3b+EAM descriptors.
The low-dimensional 2b+3b+EAM GAP is for speedup reasons tabulated by mapping the energy predictions of each term onto suitable grids. 
The pair term (2b) is tabulated as a function of distance, the three-body term (3b) on a grid of $[r_{ij}, r_{ik}, \cos{(\theta_{ijk})}]$ points, and the EAM term becomes a traditional EAM potential file where the pairwise density is tabulated as a function of distance and the embedding energy as a function of EAM density. 
The final energy and force are evaluated with cubic spline interpolations for each term (one-dimensional spline for the pairwise and EAM terms, and three-dimensional spline for the three-body term). 
More details can be found in our previous works~~\cite{byggmastar2022multiscale, byggmastar2022simple} and in the Supplementary Appendix C.
The trainings of the ML-GAPs were done with the QUIP package~\cite{quip, gap2010}. 
The phonon dispersion calculations were performed using the Phonopy package~\cite{phonopy2015}.
The testing ring statistics analysis was done with the R.I.N.G.S. package~\cite{leroux2010ring}. 

\section*{Data Availability}

The relevant structural training data (given in EXTXYZ format), the potential parameter files, and the exemplary LAMMPS inputs are openly available in the figshare repository at \url{https://doi.org/10.6084/m9.figshare.21731426.v1}. The identifier for the soapGAP potential (given in the XML file) is GAP\_2022\_9\_22\_480\_23\_27\_2\_249.

\bibliographystyle{apsrev4-2}
\bibliography{final.bib}

\begin{thebibliography}{79}%
\makeatletter
\providecommand \@ifxundefined [1]{%
 \@ifx{#1\undefined}
}%
\providecommand \@ifnum [1]{%
 \ifnum #1\expandafter \@firstoftwo
 \else \expandafter \@secondoftwo
 \fi
}%
\providecommand \@ifx [1]{%
 \ifx #1\expandafter \@firstoftwo
 \else \expandafter \@secondoftwo
 \fi
}%
\providecommand \natexlab [1]{#1}%
\providecommand \enquote  [1]{``#1''}%
\providecommand \bibnamefont  [1]{#1}%
\providecommand \bibfnamefont [1]{#1}%
\providecommand \citenamefont [1]{#1}%
\providecommand \href@noop [0]{\@secondoftwo}%
\providecommand \href [0]{\begingroup \@sanitize@url \@href}%
\providecommand \@href[1]{\@@startlink{#1}\@@href}%
\providecommand \@@href[1]{\endgroup#1\@@endlink}%
\providecommand \@sanitize@url [0]{\catcode `\\12\catcode `\$12\catcode
  `\&12\catcode `\#12\catcode `\^12\catcode `\_12\catcode `\%12\relax}%
\providecommand \@@startlink[1]{}%
\providecommand \@@endlink[0]{}%
\providecommand \url  [0]{\begingroup\@sanitize@url \@url }%
\providecommand \@url [1]{\endgroup\@href {#1}{\urlprefix }}%
\providecommand \urlprefix  [0]{URL }%
\providecommand \Eprint [0]{\href }%
\providecommand \doibase [0]{https://doi.org/}%
\providecommand \selectlanguage [0]{\@gobble}%
\providecommand \bibinfo  [0]{\@secondoftwo}%
\providecommand \bibfield  [0]{\@secondoftwo}%
\providecommand \translation [1]{[#1]}%
\providecommand \BibitemOpen [0]{}%
\providecommand \bibitemStop [0]{}%
\providecommand \bibitemNoStop [0]{.\EOS\space}%
\providecommand \EOS [0]{\spacefactor3000\relax}%
\providecommand \BibitemShut  [1]{\csname bibitem#1\endcsname}%
\let\auto@bib@innerbib\@empty
%</preamble>
\bibitem [{\citenamefont {Pearton}\ \emph
  {et~al.}(2018{\natexlab{a}})\citenamefont {Pearton}, \citenamefont {Yang},
  \citenamefont {Cary}, \citenamefont {Ren}, \citenamefont {Kim}, \citenamefont
  {Tadjer},\ and\ \citenamefont {Mastro}}]{pearton2018a}%
  \BibitemOpen
  \bibfield  {author} {\bibinfo {author} {\bibfnamefont {S.~J.}\ \bibnamefont
  {Pearton}}, \bibinfo {author} {\bibfnamefont {J.}~\bibnamefont {Yang}},
  \bibinfo {author} {\bibfnamefont {P.~H.}\ \bibnamefont {Cary}}, \bibinfo
  {author} {\bibfnamefont {F.}~\bibnamefont {Ren}}, \bibinfo {author}
  {\bibfnamefont {J.~H.}\ \bibnamefont {Kim}}, \bibinfo {author} {\bibfnamefont
  {M.~J.}\ \bibnamefont {Tadjer}},\ and\ \bibinfo {author} {\bibfnamefont
  {M.~A.}\ \bibnamefont {Mastro}},\ }\href {https://doi.org/10.1063/1.5006941}
  {\bibfield  {journal} {\bibinfo  {journal} {Appl. Phys. Rev.}\ }\textbf
  {\bibinfo {volume} {5}},\ \bibinfo {pages} {011301} (\bibinfo {year}
  {2018}{\natexlab{a}})}\BibitemShut {NoStop}%
\bibitem [{\citenamefont {Jesenovec}\ \emph {et~al.}(2022)\citenamefont
  {Jesenovec}, \citenamefont {Pansegrau}, \citenamefont {McCluskey},
  \citenamefont {McCloy}, \citenamefont {Gustafson}, \citenamefont
  {Halliburton},\ and\ \citenamefont {Varley}}]{jesenovec2022persistent}%
  \BibitemOpen
  \bibfield  {author} {\bibinfo {author} {\bibfnamefont {J.}~\bibnamefont
  {Jesenovec}}, \bibinfo {author} {\bibfnamefont {C.}~\bibnamefont
  {Pansegrau}}, \bibinfo {author} {\bibfnamefont {M.~D.}\ \bibnamefont
  {McCluskey}}, \bibinfo {author} {\bibfnamefont {J.~S.}\ \bibnamefont
  {McCloy}}, \bibinfo {author} {\bibfnamefont {T.~D.}\ \bibnamefont
  {Gustafson}}, \bibinfo {author} {\bibfnamefont {L.~E.}\ \bibnamefont
  {Halliburton}},\ and\ \bibinfo {author} {\bibfnamefont {J.~B.}\ \bibnamefont
  {Varley}},\ }\href {https://doi.org/10.1103/PhysRevLett.128.077402}
  {\bibfield  {journal} {\bibinfo  {journal} {Phys. Rev. Lett.}\ }\textbf
  {\bibinfo {volume} {128}},\ \bibinfo {pages} {077402} (\bibinfo {year}
  {2022})}\BibitemShut {NoStop}%
\bibitem [{\citenamefont {Pearton}\ \emph
  {et~al.}(2018{\natexlab{b}})\citenamefont {Pearton}, \citenamefont {Ren},
  \citenamefont {Tadjer},\ and\ \citenamefont {Kim}}]{pearton2018perspective}%
  \BibitemOpen
  \bibfield  {author} {\bibinfo {author} {\bibfnamefont {S.~J.}\ \bibnamefont
  {Pearton}}, \bibinfo {author} {\bibfnamefont {F.}~\bibnamefont {Ren}},
  \bibinfo {author} {\bibfnamefont {M.}~\bibnamefont {Tadjer}},\ and\ \bibinfo
  {author} {\bibfnamefont {J.}~\bibnamefont {Kim}},\ }\href
  {https://doi.org/10.1063/1.5062841} {\bibfield  {journal} {\bibinfo
  {journal} {J. Appl. Phys.}\ }\textbf {\bibinfo {volume} {124}},\ \bibinfo
  {pages} {220901} (\bibinfo {year} {2018}{\natexlab{b}})}\BibitemShut
  {NoStop}%
\bibitem [{\citenamefont {Zhang}\ \emph {et~al.}(2022)\citenamefont {Zhang},
  \citenamefont {Dong}, \citenamefont {Dang}, \citenamefont {Zhang},
  \citenamefont {Yan}, \citenamefont {Xiang}, \citenamefont {Su}, \citenamefont
  {Liu}, \citenamefont {Si}, \citenamefont {Gao}, \citenamefont {Kong},
  \citenamefont {Zhou},\ and\ \citenamefont {Hao}}]{zhang2022ultra}%
  \BibitemOpen
  \bibfield  {author} {\bibinfo {author} {\bibfnamefont {J.}~\bibnamefont
  {Zhang}}, \bibinfo {author} {\bibfnamefont {P.}~\bibnamefont {Dong}},
  \bibinfo {author} {\bibfnamefont {K.}~\bibnamefont {Dang}}, \bibinfo {author}
  {\bibfnamefont {Y.}~\bibnamefont {Zhang}}, \bibinfo {author} {\bibfnamefont
  {Q.}~\bibnamefont {Yan}}, \bibinfo {author} {\bibfnamefont {H.}~\bibnamefont
  {Xiang}}, \bibinfo {author} {\bibfnamefont {J.}~\bibnamefont {Su}}, \bibinfo
  {author} {\bibfnamefont {Z.}~\bibnamefont {Liu}}, \bibinfo {author}
  {\bibfnamefont {M.}~\bibnamefont {Si}}, \bibinfo {author} {\bibfnamefont
  {J.}~\bibnamefont {Gao}}, \bibinfo {author} {\bibfnamefont {M.}~\bibnamefont
  {Kong}}, \bibinfo {author} {\bibfnamefont {H.}~\bibnamefont {Zhou}},\ and\
  \bibinfo {author} {\bibfnamefont {Y.}~\bibnamefont {Hao}},\ }\href
  {https://doi.org/10.1038/s41467-022-31664-y} {\bibfield  {journal} {\bibinfo
  {journal} {Nat. Commun.}\ }\textbf {\bibinfo {volume} {13}},\ \bibinfo
  {pages} {3900} (\bibinfo {year} {2022})}\BibitemShut {NoStop}%
\bibitem [{\citenamefont {Kim}\ and\ \citenamefont
  {Kim}(2020)}]{kim2020highly}%
  \BibitemOpen
  \bibfield  {author} {\bibinfo {author} {\bibfnamefont {S.}~\bibnamefont
  {Kim}}\ and\ \bibinfo {author} {\bibfnamefont {J.}~\bibnamefont {Kim}},\
  }\href {https://doi.org/10.1063/5.0030400} {\bibfield  {journal} {\bibinfo
  {journal} {Appl. Phys. Lett.}\ }\textbf {\bibinfo {volume} {117}},\ \bibinfo
  {pages} {261101} (\bibinfo {year} {2020})}\BibitemShut {NoStop}%
\bibitem [{\citenamefont {Wang}\ \emph {et~al.}(2021)\citenamefont {Wang},
  \citenamefont {Li}, \citenamefont {Cao}, \citenamefont {Shen}, \citenamefont
  {Zhang}, \citenamefont {Yang}, \citenamefont {Dong}, \citenamefont {Zhou},
  \citenamefont {Zhang}, \citenamefont {Tang},\ and\ \citenamefont
  {Wu}}]{wang2021ultrahigh}%
  \BibitemOpen
  \bibfield  {author} {\bibinfo {author} {\bibfnamefont {Y.}~\bibnamefont
  {Wang}}, \bibinfo {author} {\bibfnamefont {H.}~\bibnamefont {Li}}, \bibinfo
  {author} {\bibfnamefont {J.}~\bibnamefont {Cao}}, \bibinfo {author}
  {\bibfnamefont {J.}~\bibnamefont {Shen}}, \bibinfo {author} {\bibfnamefont
  {Q.}~\bibnamefont {Zhang}}, \bibinfo {author} {\bibfnamefont
  {Y.}~\bibnamefont {Yang}}, \bibinfo {author} {\bibfnamefont {Z.}~\bibnamefont
  {Dong}}, \bibinfo {author} {\bibfnamefont {T.}~\bibnamefont {Zhou}}, \bibinfo
  {author} {\bibfnamefont {Y.}~\bibnamefont {Zhang}}, \bibinfo {author}
  {\bibfnamefont {W.}~\bibnamefont {Tang}},\ and\ \bibinfo {author}
  {\bibfnamefont {Z.}~\bibnamefont {Wu}},\ }\href
  {https://doi.org/10.1021/acsnano.1c06567} {\bibfield  {journal} {\bibinfo
  {journal} {ACS Nano}\ }\textbf {\bibinfo {volume} {15}},\ \bibinfo {pages}
  {16654} (\bibinfo {year} {2021})}\BibitemShut {NoStop}%
\bibitem [{\citenamefont {Tang}\ \emph {et~al.}(2022)\citenamefont {Tang},
  \citenamefont {Li}, \citenamefont {Zhao}, \citenamefont {Sui}, \citenamefont
  {Liang}, \citenamefont {Liu}, \citenamefont {Liao}, \citenamefont {Babatain},
  \citenamefont {Lin}, \citenamefont {Wang}, \citenamefont {Lu}, \citenamefont
  {Alqatari}, \citenamefont {Mei}, \citenamefont {Tang},\ and\ \citenamefont
  {Li}}]{tang2022quasiepitaxial}%
  \BibitemOpen
  \bibfield  {author} {\bibinfo {author} {\bibfnamefont {X.}~\bibnamefont
  {Tang}}, \bibinfo {author} {\bibfnamefont {K.-H.}\ \bibnamefont {Li}},
  \bibinfo {author} {\bibfnamefont {Y.}~\bibnamefont {Zhao}}, \bibinfo {author}
  {\bibfnamefont {Y.}~\bibnamefont {Sui}}, \bibinfo {author} {\bibfnamefont
  {H.}~\bibnamefont {Liang}}, \bibinfo {author} {\bibfnamefont
  {Z.}~\bibnamefont {Liu}}, \bibinfo {author} {\bibfnamefont {C.-H.}\
  \bibnamefont {Liao}}, \bibinfo {author} {\bibfnamefont {W.}~\bibnamefont
  {Babatain}}, \bibinfo {author} {\bibfnamefont {R.}~\bibnamefont {Lin}},
  \bibinfo {author} {\bibfnamefont {C.}~\bibnamefont {Wang}}, \bibinfo {author}
  {\bibfnamefont {Y.}~\bibnamefont {Lu}}, \bibinfo {author} {\bibfnamefont
  {F.~S.}\ \bibnamefont {Alqatari}}, \bibinfo {author} {\bibfnamefont
  {Z.}~\bibnamefont {Mei}}, \bibinfo {author} {\bibfnamefont {W.}~\bibnamefont
  {Tang}},\ and\ \bibinfo {author} {\bibfnamefont {X.}~\bibnamefont {Li}},\
  }\href {https://doi.org/10.1021/acsami.1c15560} {\bibfield  {journal}
  {\bibinfo  {journal} {ACS Appl. Mater. Interfaces}\ }\textbf {\bibinfo
  {volume} {14}},\ \bibinfo {pages} {1304} (\bibinfo {year}
  {2022})}\BibitemShut {NoStop}%
\bibitem [{\citenamefont {Mazeina}\ \emph {et~al.}(2010)\citenamefont
  {Mazeina}, \citenamefont {Perkins}, \citenamefont {Bermudez}, \citenamefont
  {Arnold},\ and\ \citenamefont {Prokes}}]{mazeina2010functionalized}%
  \BibitemOpen
  \bibfield  {author} {\bibinfo {author} {\bibfnamefont {L.}~\bibnamefont
  {Mazeina}}, \bibinfo {author} {\bibfnamefont {F.~K.}\ \bibnamefont
  {Perkins}}, \bibinfo {author} {\bibfnamefont {V.~M.}\ \bibnamefont
  {Bermudez}}, \bibinfo {author} {\bibfnamefont {S.~P.}\ \bibnamefont
  {Arnold}},\ and\ \bibinfo {author} {\bibfnamefont {S.~M.}\ \bibnamefont
  {Prokes}},\ }\href {https://doi.org/10.1021/la101760k} {\bibfield  {journal}
  {\bibinfo  {journal} {Langmuir}\ }\textbf {\bibinfo {volume} {26}},\ \bibinfo
  {pages} {13722} (\bibinfo {year} {2010})}\BibitemShut {NoStop}%
\bibitem [{\citenamefont {Zhao}\ \emph
  {et~al.}(2021{\natexlab{a}})\citenamefont {Zhao}, \citenamefont {Huang},
  \citenamefont {Yin}, \citenamefont {Liao}, \citenamefont {Mo}, \citenamefont
  {Qian}, \citenamefont {Guo}, \citenamefont {Chen}, \citenamefont {Zhang},\
  and\ \citenamefont {Hua}}]{SFzhao2021two}%
  \BibitemOpen
  \bibfield  {author} {\bibinfo {author} {\bibfnamefont {J.}~\bibnamefont
  {Zhao}}, \bibinfo {author} {\bibfnamefont {X.}~\bibnamefont {Huang}},
  \bibinfo {author} {\bibfnamefont {Y.}~\bibnamefont {Yin}}, \bibinfo {author}
  {\bibfnamefont {Y.}~\bibnamefont {Liao}}, \bibinfo {author} {\bibfnamefont
  {H.}~\bibnamefont {Mo}}, \bibinfo {author} {\bibfnamefont {Q.}~\bibnamefont
  {Qian}}, \bibinfo {author} {\bibfnamefont {Y.}~\bibnamefont {Guo}}, \bibinfo
  {author} {\bibfnamefont {X.}~\bibnamefont {Chen}}, \bibinfo {author}
  {\bibfnamefont {Z.}~\bibnamefont {Zhang}},\ and\ \bibinfo {author}
  {\bibfnamefont {M.}~\bibnamefont {Hua}},\ }\href
  {https://doi.org/10.1021/acs.jpclett.1c01393} {\bibfield  {journal} {\bibinfo
   {journal} {J. Phys. Chem. Lett.}\ }\textbf {\bibinfo {volume} {12}},\
  \bibinfo {pages} {5813} (\bibinfo {year} {2021}{\natexlab{a}})}\BibitemShut
  {NoStop}%
\bibitem [{\citenamefont {Zavabeti}\ \emph {et~al.}(2017)\citenamefont
  {Zavabeti}, \citenamefont {Ou}, \citenamefont {Carey}, \citenamefont {Syed},
  \citenamefont {Orrell-Trigg}, \citenamefont {Mayes}, \citenamefont {Xu},
  \citenamefont {Kavehei}, \citenamefont {O{\textquoteright}Mullane},
  \citenamefont {Kaner}, \citenamefont {Kalantar-zadeh},\ and\ \citenamefont
  {Daeneke}}]{zavabeti2017liquid}%
  \BibitemOpen
  \bibfield  {author} {\bibinfo {author} {\bibfnamefont {A.}~\bibnamefont
  {Zavabeti}}, \bibinfo {author} {\bibfnamefont {J.~Z.}\ \bibnamefont {Ou}},
  \bibinfo {author} {\bibfnamefont {B.~J.}\ \bibnamefont {Carey}}, \bibinfo
  {author} {\bibfnamefont {N.}~\bibnamefont {Syed}}, \bibinfo {author}
  {\bibfnamefont {R.}~\bibnamefont {Orrell-Trigg}}, \bibinfo {author}
  {\bibfnamefont {E.~L.~H.}\ \bibnamefont {Mayes}}, \bibinfo {author}
  {\bibfnamefont {C.}~\bibnamefont {Xu}}, \bibinfo {author} {\bibfnamefont
  {O.}~\bibnamefont {Kavehei}}, \bibinfo {author} {\bibfnamefont {A.~P.}\
  \bibnamefont {O{\textquoteright}Mullane}}, \bibinfo {author} {\bibfnamefont
  {R.~B.}\ \bibnamefont {Kaner}}, \bibinfo {author} {\bibfnamefont
  {K.}~\bibnamefont {Kalantar-zadeh}},\ and\ \bibinfo {author} {\bibfnamefont
  {T.}~\bibnamefont {Daeneke}},\ }\href
  {https://doi.org/10.1126/science.aao4249} {\bibfield  {journal} {\bibinfo
  {journal} {Science}\ }\textbf {\bibinfo {volume} {358}},\ \bibinfo {pages}
  {332} (\bibinfo {year} {2017})}\BibitemShut {NoStop}%
\bibitem [{\citenamefont {Wurdack}\ \emph {et~al.}(2021)\citenamefont
  {Wurdack}, \citenamefont {Yun}, \citenamefont {Estrecho}, \citenamefont
  {Syed}, \citenamefont {Bhattacharyya}, \citenamefont {Pieczarka},
  \citenamefont {Zavabeti}, \citenamefont {Chen}, \citenamefont {Haas},
  \citenamefont {M{\"u}ller}, \citenamefont {Lockrey}, \citenamefont {Bao},
  \citenamefont {Schneider}, \citenamefont {Lu}, \citenamefont {Fuhrer},
  \citenamefont {Truscott}, \citenamefont {Daeneke},\ and\ \citenamefont
  {Ostrovskaya}}]{wurdack2021ultrathin}%
  \BibitemOpen
  \bibfield  {author} {\bibinfo {author} {\bibfnamefont {M.}~\bibnamefont
  {Wurdack}}, \bibinfo {author} {\bibfnamefont {T.}~\bibnamefont {Yun}},
  \bibinfo {author} {\bibfnamefont {E.}~\bibnamefont {Estrecho}}, \bibinfo
  {author} {\bibfnamefont {N.}~\bibnamefont {Syed}}, \bibinfo {author}
  {\bibfnamefont {S.}~\bibnamefont {Bhattacharyya}}, \bibinfo {author}
  {\bibfnamefont {M.}~\bibnamefont {Pieczarka}}, \bibinfo {author}
  {\bibfnamefont {A.}~\bibnamefont {Zavabeti}}, \bibinfo {author}
  {\bibfnamefont {S.}~\bibnamefont {Chen}}, \bibinfo {author} {\bibfnamefont
  {B.}~\bibnamefont {Haas}}, \bibinfo {author} {\bibfnamefont {J.}~\bibnamefont
  {M{\"u}ller}}, \bibinfo {author} {\bibfnamefont {M.~N.}\ \bibnamefont
  {Lockrey}}, \bibinfo {author} {\bibfnamefont {Q.}~\bibnamefont {Bao}},
  \bibinfo {author} {\bibfnamefont {C.}~\bibnamefont {Schneider}}, \bibinfo
  {author} {\bibfnamefont {Y.}~\bibnamefont {Lu}}, \bibinfo {author}
  {\bibfnamefont {M.~S.}\ \bibnamefont {Fuhrer}}, \bibinfo {author}
  {\bibfnamefont {A.~G.}\ \bibnamefont {Truscott}}, \bibinfo {author}
  {\bibfnamefont {T.}~\bibnamefont {Daeneke}},\ and\ \bibinfo {author}
  {\bibfnamefont {E.~A.}\ \bibnamefont {Ostrovskaya}},\ }\href
  {https://doi.org/10.1002/adma.202005732} {\bibfield  {journal} {\bibinfo
  {journal} {Adv. Mater.}\ }\textbf {\bibinfo {volume} {33}},\ \bibinfo {pages}
  {2005732} (\bibinfo {year} {2021})}\BibitemShut {NoStop}%
\bibitem [{\citenamefont {Zhao}\ \emph {et~al.}(2022)\citenamefont {Zhao},
  \citenamefont {Wang}, \citenamefont {Chen}, \citenamefont {Zhang},\ and\
  \citenamefont {Hua}}]{SFzhao2022bilayer}%
  \BibitemOpen
  \bibfield  {author} {\bibinfo {author} {\bibfnamefont {J.}~\bibnamefont
  {Zhao}}, \bibinfo {author} {\bibfnamefont {X.}~\bibnamefont {Wang}}, \bibinfo
  {author} {\bibfnamefont {H.}~\bibnamefont {Chen}}, \bibinfo {author}
  {\bibfnamefont {Z.}~\bibnamefont {Zhang}},\ and\ \bibinfo {author}
  {\bibfnamefont {M.}~\bibnamefont {Hua}},\ }\href
  {https://doi.org/10.1021/acs.chemmater.1c04245} {\bibfield  {journal}
  {\bibinfo  {journal} {Chem. Mater.}\ }\textbf {\bibinfo {volume} {34}},\
  \bibinfo {pages} {3648} (\bibinfo {year} {2022})}\BibitemShut {NoStop}%
\bibitem [{\citenamefont {Playford}\ \emph {et~al.}(2013)\citenamefont
  {Playford}, \citenamefont {Hannon}, \citenamefont {Barney},\ and\
  \citenamefont {Walton}}]{playford2013structures}%
  \BibitemOpen
  \bibfield  {author} {\bibinfo {author} {\bibfnamefont {H.~Y.}\ \bibnamefont
  {Playford}}, \bibinfo {author} {\bibfnamefont {A.~C.}\ \bibnamefont
  {Hannon}}, \bibinfo {author} {\bibfnamefont {E.~R.}\ \bibnamefont {Barney}},\
  and\ \bibinfo {author} {\bibfnamefont {R.~I.}\ \bibnamefont {Walton}},\
  }\href {https://doi.org/https://doi.org/10.1002/chem.201203359} {\bibfield
  {journal} {\bibinfo  {journal} {Chem. - Eur. J.}\ }\textbf {\bibinfo {volume}
  {19}},\ \bibinfo {pages} {2803} (\bibinfo {year} {2013})}\BibitemShut
  {NoStop}%
\bibitem [{\citenamefont {Cora}\ \emph {et~al.}(2017)\citenamefont {Cora},
  \citenamefont {Mezzadri}, \citenamefont {Boschi}, \citenamefont {Bosi},
  \citenamefont {{\v C}aplovi{\v c}ov{\'a}}, \citenamefont {Calestani},
  \citenamefont {D{\'o}dony}, \citenamefont {P{\'e}cz},\ and\ \citenamefont
  {Fornari}}]{cora2017real}%
  \BibitemOpen
  \bibfield  {author} {\bibinfo {author} {\bibfnamefont {I.}~\bibnamefont
  {Cora}}, \bibinfo {author} {\bibfnamefont {F.}~\bibnamefont {Mezzadri}},
  \bibinfo {author} {\bibfnamefont {F.}~\bibnamefont {Boschi}}, \bibinfo
  {author} {\bibfnamefont {M.}~\bibnamefont {Bosi}}, \bibinfo {author}
  {\bibfnamefont {M.}~\bibnamefont {{\v C}aplovi{\v c}ov{\'a}}}, \bibinfo
  {author} {\bibfnamefont {G.}~\bibnamefont {Calestani}}, \bibinfo {author}
  {\bibfnamefont {I.}~\bibnamefont {D{\'o}dony}}, \bibinfo {author}
  {\bibfnamefont {B.}~\bibnamefont {P{\'e}cz}},\ and\ \bibinfo {author}
  {\bibfnamefont {R.}~\bibnamefont {Fornari}},\ }\href
  {https://doi.org/10.1039/C7CE00123A} {\bibfield  {journal} {\bibinfo
  {journal} {CrystEngComm}\ }\textbf {\bibinfo {volume} {19}},\ \bibinfo
  {pages} {1509} (\bibinfo {year} {2017})}\BibitemShut {NoStop}%
\bibitem [{\citenamefont {Swallow}\ \emph {et~al.}(2020)\citenamefont
  {Swallow}, \citenamefont {Vorwerk}, \citenamefont {Mazzolini}, \citenamefont
  {Vogt}, \citenamefont {Bierwagen}, \citenamefont {Karg}, \citenamefont
  {Eickhoff}, \citenamefont {Sch{\"o}rmann}, \citenamefont {Wagner},
  \citenamefont {Roberts}, \citenamefont {Chalker}, \citenamefont {Smiles},
  \citenamefont {Murgatroyd}, \citenamefont {Razek}, \citenamefont
  {Lebens-Higgins}, \citenamefont {Piper}, \citenamefont {Jones}, \citenamefont
  {Thakur}, \citenamefont {Lee}, \citenamefont {Varley}, \citenamefont
  {Furthm{\"u}ller}, \citenamefont {Draxl}, \citenamefont {Veal},\ and\
  \citenamefont {Regoutz}}]{swallow2020influence}%
  \BibitemOpen
  \bibfield  {author} {\bibinfo {author} {\bibfnamefont {J.~E.~N.}\
  \bibnamefont {Swallow}}, \bibinfo {author} {\bibfnamefont {C.}~\bibnamefont
  {Vorwerk}}, \bibinfo {author} {\bibfnamefont {P.}~\bibnamefont {Mazzolini}},
  \bibinfo {author} {\bibfnamefont {P.}~\bibnamefont {Vogt}}, \bibinfo {author}
  {\bibfnamefont {O.}~\bibnamefont {Bierwagen}}, \bibinfo {author}
  {\bibfnamefont {A.}~\bibnamefont {Karg}}, \bibinfo {author} {\bibfnamefont
  {M.}~\bibnamefont {Eickhoff}}, \bibinfo {author} {\bibfnamefont
  {J.}~\bibnamefont {Sch{\"o}rmann}}, \bibinfo {author} {\bibfnamefont {M.~R.}\
  \bibnamefont {Wagner}}, \bibinfo {author} {\bibfnamefont {J.~W.}\
  \bibnamefont {Roberts}}, \bibinfo {author} {\bibfnamefont {P.~R.}\
  \bibnamefont {Chalker}}, \bibinfo {author} {\bibfnamefont {M.~J.}\
  \bibnamefont {Smiles}}, \bibinfo {author} {\bibfnamefont {P.}~\bibnamefont
  {Murgatroyd}}, \bibinfo {author} {\bibfnamefont {S.~A.}\ \bibnamefont
  {Razek}}, \bibinfo {author} {\bibfnamefont {Z.~W.}\ \bibnamefont
  {Lebens-Higgins}}, \bibinfo {author} {\bibfnamefont {L.~F.~J.}\ \bibnamefont
  {Piper}}, \bibinfo {author} {\bibfnamefont {L.~A.~H.}\ \bibnamefont {Jones}},
  \bibinfo {author} {\bibfnamefont {P.~K.}\ \bibnamefont {Thakur}}, \bibinfo
  {author} {\bibfnamefont {T.-L.}\ \bibnamefont {Lee}}, \bibinfo {author}
  {\bibfnamefont {J.~B.}\ \bibnamefont {Varley}}, \bibinfo {author}
  {\bibfnamefont {J.}~\bibnamefont {Furthm{\"u}ller}}, \bibinfo {author}
  {\bibfnamefont {C.}~\bibnamefont {Draxl}}, \bibinfo {author} {\bibfnamefont
  {T.~D.}\ \bibnamefont {Veal}},\ and\ \bibinfo {author} {\bibfnamefont
  {A.}~\bibnamefont {Regoutz}},\ }\href
  {https://doi.org/10.1021/acs.chemmater.0c02465} {\bibfield  {journal}
  {\bibinfo  {journal} {Chem. Mater.}\ }\textbf {\bibinfo {volume} {32}},\
  \bibinfo {pages} {8460} (\bibinfo {year} {2020})}\BibitemShut {NoStop}%
\bibitem [{\citenamefont {Mu}\ and\ \citenamefont {Van~de
  Walle}(2022)}]{mu2022phase}%
  \BibitemOpen
  \bibfield  {author} {\bibinfo {author} {\bibfnamefont {S.}~\bibnamefont
  {Mu}}\ and\ \bibinfo {author} {\bibfnamefont {C.~G.}\ \bibnamefont {Van~de
  Walle}},\ }\href {https://doi.org/10.1103/PhysRevMaterials.6.104601}
  {\bibfield  {journal} {\bibinfo  {journal} {Phys. Rev. Mater.}\ }\textbf
  {\bibinfo {volume} {6}},\ \bibinfo {pages} {104601} (\bibinfo {year}
  {2022})}\BibitemShut {NoStop}%
\bibitem [{\citenamefont {Remeika}\ and\ \citenamefont
  {Marezio}(1966)}]{remeika1966growth}%
  \BibitemOpen
  \bibfield  {author} {\bibinfo {author} {\bibfnamefont {J.~P.}\ \bibnamefont
  {Remeika}}\ and\ \bibinfo {author} {\bibfnamefont {M.}~\bibnamefont
  {Marezio}},\ }\href {https://doi.org/10.1063/1.1754500} {\bibfield  {journal}
  {\bibinfo  {journal} {Appl. Phys. Lett.}\ }\textbf {\bibinfo {volume} {8}},\
  \bibinfo {pages} {87} (\bibinfo {year} {1966})}\BibitemShut {NoStop}%
\bibitem [{\citenamefont {Lion}\ \emph {et~al.}(2022)\citenamefont {Lion},
  \citenamefont {Pavone},\ and\ \citenamefont {Draxl}}]{lion2022elastic}%
  \BibitemOpen
  \bibfield  {author} {\bibinfo {author} {\bibfnamefont {K.}~\bibnamefont
  {Lion}}, \bibinfo {author} {\bibfnamefont {P.}~\bibnamefont {Pavone}},\ and\
  \bibinfo {author} {\bibfnamefont {C.}~\bibnamefont {Draxl}},\ }\href
  {https://doi.org/10.1103/PhysRevMaterials.6.013601} {\bibfield  {journal}
  {\bibinfo  {journal} {Phys. Rev. Mater.}\ }\textbf {\bibinfo {volume} {6}},\
  \bibinfo {pages} {013601} (\bibinfo {year} {2022})}\BibitemShut {NoStop}%
\bibitem [{\citenamefont {Sun}\ \emph {et~al.}(2018)\citenamefont {Sun},
  \citenamefont {Li}, \citenamefont {Castanedo}, \citenamefont {Okur},
  \citenamefont {Tompa}, \citenamefont {Salagaj}, \citenamefont {Lopatin},
  \citenamefont {Genovese},\ and\ \citenamefont {Li}}]{sun2018hcl}%
  \BibitemOpen
  \bibfield  {author} {\bibinfo {author} {\bibfnamefont {H.}~\bibnamefont
  {Sun}}, \bibinfo {author} {\bibfnamefont {K.-H.}\ \bibnamefont {Li}},
  \bibinfo {author} {\bibfnamefont {C.~G.~T.}\ \bibnamefont {Castanedo}},
  \bibinfo {author} {\bibfnamefont {S.}~\bibnamefont {Okur}}, \bibinfo {author}
  {\bibfnamefont {G.~S.}\ \bibnamefont {Tompa}}, \bibinfo {author}
  {\bibfnamefont {T.}~\bibnamefont {Salagaj}}, \bibinfo {author} {\bibfnamefont
  {S.}~\bibnamefont {Lopatin}}, \bibinfo {author} {\bibfnamefont
  {A.}~\bibnamefont {Genovese}},\ and\ \bibinfo {author} {\bibfnamefont
  {X.}~\bibnamefont {Li}},\ }\href {https://doi.org/10.1021/acs.cgd.7b01791}
  {\bibfield  {journal} {\bibinfo  {journal} {Cryst. Growth Des.}\ }\textbf
  {\bibinfo {volume} {18}},\ \bibinfo {pages} {2370} (\bibinfo {year}
  {2018})}\BibitemShut {NoStop}%
\bibitem [{\citenamefont {Wheeler}\ \emph {et~al.}(2020)\citenamefont
  {Wheeler}, \citenamefont {Nepal}, \citenamefont {Boris}, \citenamefont
  {Qadri}, \citenamefont {Nyakiti}, \citenamefont {Lang}, \citenamefont
  {Koehler}, \citenamefont {Foster}, \citenamefont {Walton}, \citenamefont
  {Eddy},\ and\ \citenamefont {Meyer}}]{wheeler2020phase}%
  \BibitemOpen
  \bibfield  {author} {\bibinfo {author} {\bibfnamefont {V.~D.}\ \bibnamefont
  {Wheeler}}, \bibinfo {author} {\bibfnamefont {N.}~\bibnamefont {Nepal}},
  \bibinfo {author} {\bibfnamefont {D.~R.}\ \bibnamefont {Boris}}, \bibinfo
  {author} {\bibfnamefont {S.~B.}\ \bibnamefont {Qadri}}, \bibinfo {author}
  {\bibfnamefont {L.~O.}\ \bibnamefont {Nyakiti}}, \bibinfo {author}
  {\bibfnamefont {A.}~\bibnamefont {Lang}}, \bibinfo {author} {\bibfnamefont
  {A.}~\bibnamefont {Koehler}}, \bibinfo {author} {\bibfnamefont
  {G.}~\bibnamefont {Foster}}, \bibinfo {author} {\bibfnamefont {S.~G.}\
  \bibnamefont {Walton}}, \bibinfo {author} {\bibfnamefont {C.~R.}\
  \bibnamefont {Eddy}},\ and\ \bibinfo {author} {\bibfnamefont {D.~J.}\
  \bibnamefont {Meyer}},\ }\href
  {https://doi.org/10.1021/acs.chemmater.9b03926} {\bibfield  {journal}
  {\bibinfo  {journal} {Chem. Mater.}\ }\textbf {\bibinfo {volume} {32}},\
  \bibinfo {pages} {1140} (\bibinfo {year} {2020})}\BibitemShut {NoStop}%
\bibitem [{\citenamefont {He}\ \emph {et~al.}(2006)\citenamefont {He},
  \citenamefont {Orlando}, \citenamefont {Blanco}, \citenamefont {Pandey},
  \citenamefont {Amzallag}, \citenamefont {Baraille},\ and\ \citenamefont
  {R\'erat}}]{he2006first}%
  \BibitemOpen
  \bibfield  {author} {\bibinfo {author} {\bibfnamefont {H.}~\bibnamefont
  {He}}, \bibinfo {author} {\bibfnamefont {R.}~\bibnamefont {Orlando}},
  \bibinfo {author} {\bibfnamefont {M.~A.}\ \bibnamefont {Blanco}}, \bibinfo
  {author} {\bibfnamefont {R.}~\bibnamefont {Pandey}}, \bibinfo {author}
  {\bibfnamefont {E.}~\bibnamefont {Amzallag}}, \bibinfo {author}
  {\bibfnamefont {I.}~\bibnamefont {Baraille}},\ and\ \bibinfo {author}
  {\bibfnamefont {M.}~\bibnamefont {R\'erat}},\ }\href
  {https://doi.org/10.1103/PhysRevB.74.195123} {\bibfield  {journal} {\bibinfo
  {journal} {Phys. Rev. B}\ }\textbf {\bibinfo {volume} {74}},\ \bibinfo
  {pages} {195123} (\bibinfo {year} {2006})}\BibitemShut {NoStop}%
\bibitem [{\citenamefont {Uno}\ \emph {et~al.}(2020)\citenamefont {Uno},
  \citenamefont {Ohta},\ and\ \citenamefont {Tanaka}}]{uno2020growth}%
  \BibitemOpen
  \bibfield  {author} {\bibinfo {author} {\bibfnamefont {K.}~\bibnamefont
  {Uno}}, \bibinfo {author} {\bibfnamefont {M.}~\bibnamefont {Ohta}},\ and\
  \bibinfo {author} {\bibfnamefont {I.}~\bibnamefont {Tanaka}},\ }\href
  {https://doi.org/10.1063/5.0014056} {\bibfield  {journal} {\bibinfo
  {journal} {Appl. Phys. Lett.}\ }\textbf {\bibinfo {volume} {117}},\ \bibinfo
  {pages} {052106} (\bibinfo {year} {2020})}\BibitemShut {NoStop}%
\bibitem [{\citenamefont {Mezzadri}\ \emph {et~al.}(2016)\citenamefont
  {Mezzadri}, \citenamefont {Calestani}, \citenamefont {Boschi}, \citenamefont
  {Delmonte}, \citenamefont {Bosi},\ and\ \citenamefont
  {Fornari}}]{mezzadri2016crystal}%
  \BibitemOpen
  \bibfield  {author} {\bibinfo {author} {\bibfnamefont {F.}~\bibnamefont
  {Mezzadri}}, \bibinfo {author} {\bibfnamefont {G.}~\bibnamefont {Calestani}},
  \bibinfo {author} {\bibfnamefont {F.}~\bibnamefont {Boschi}}, \bibinfo
  {author} {\bibfnamefont {D.}~\bibnamefont {Delmonte}}, \bibinfo {author}
  {\bibfnamefont {M.}~\bibnamefont {Bosi}},\ and\ \bibinfo {author}
  {\bibfnamefont {R.}~\bibnamefont {Fornari}},\ }\href
  {https://doi.org/10.1021/acs.inorgchem.6b02244} {\bibfield  {journal}
  {\bibinfo  {journal} {Inorg. Chem.}\ }\textbf {\bibinfo {volume} {55}},\
  \bibinfo {pages} {12079} (\bibinfo {year} {2016})}\BibitemShut {NoStop}%
\bibitem [{\citenamefont {Kim}\ \emph {et~al.}(2018)\citenamefont {Kim},
  \citenamefont {Tahara}, \citenamefont {Miura},\ and\ \citenamefont
  {Kim}}]{kim2018first}%
  \BibitemOpen
  \bibfield  {author} {\bibinfo {author} {\bibfnamefont {J.}~\bibnamefont
  {Kim}}, \bibinfo {author} {\bibfnamefont {D.}~\bibnamefont {Tahara}},
  \bibinfo {author} {\bibfnamefont {Y.}~\bibnamefont {Miura}},\ and\ \bibinfo
  {author} {\bibfnamefont {B.~G.}\ \bibnamefont {Kim}},\ }\href
  {https://doi.org/10.7567/apex.11.061101} {\bibfield  {journal} {\bibinfo
  {journal} {Appl. Phys. Express}\ }\textbf {\bibinfo {volume} {11}},\ \bibinfo
  {pages} {061101} (\bibinfo {year} {2018})}\BibitemShut {NoStop}%
\bibitem [{\citenamefont {Ranga}\ \emph {et~al.}(2020)\citenamefont {Ranga},
  \citenamefont {Cho}, \citenamefont {Mishra},\ and\ \citenamefont
  {Krishnamoorthy}}]{ranga2020highly}%
  \BibitemOpen
  \bibfield  {author} {\bibinfo {author} {\bibfnamefont {P.}~\bibnamefont
  {Ranga}}, \bibinfo {author} {\bibfnamefont {S.~B.}\ \bibnamefont {Cho}},
  \bibinfo {author} {\bibfnamefont {R.}~\bibnamefont {Mishra}},\ and\ \bibinfo
  {author} {\bibfnamefont {S.}~\bibnamefont {Krishnamoorthy}},\ }\href
  {https://doi.org/10.35848/1882-0786/ab9168} {\bibfield  {journal} {\bibinfo
  {journal} {Appl. Phys. Express}\ }\textbf {\bibinfo {volume} {13}},\ \bibinfo
  {pages} {061009} (\bibinfo {year} {2020})}\BibitemShut {NoStop}%
\bibitem [{\citenamefont {Xu}\ \emph {et~al.}(2019)\citenamefont {Xu},
  \citenamefont {Park}, \citenamefont {Yao}, \citenamefont {Wolverton},
  \citenamefont {Razeghi}, \citenamefont {Wu},\ and\ \citenamefont
  {Dravid}}]{xu2019strain}%
  \BibitemOpen
  \bibfield  {author} {\bibinfo {author} {\bibfnamefont {Y.}~\bibnamefont
  {Xu}}, \bibinfo {author} {\bibfnamefont {J.-H.}\ \bibnamefont {Park}},
  \bibinfo {author} {\bibfnamefont {Z.}~\bibnamefont {Yao}}, \bibinfo {author}
  {\bibfnamefont {C.}~\bibnamefont {Wolverton}}, \bibinfo {author}
  {\bibfnamefont {M.}~\bibnamefont {Razeghi}}, \bibinfo {author} {\bibfnamefont
  {J.}~\bibnamefont {Wu}},\ and\ \bibinfo {author} {\bibfnamefont {V.~P.}\
  \bibnamefont {Dravid}},\ }\href {https://doi.org/10.1021/acsami.8b17731}
  {\bibfield  {journal} {\bibinfo  {journal} {ACS Appl. Mater. Interfaces}\
  }\textbf {\bibinfo {volume} {11}},\ \bibinfo {pages} {5536} (\bibinfo {year}
  {2019})}\BibitemShut {NoStop}%
\bibitem [{\citenamefont {Cora}\ \emph {et~al.}(2020)\citenamefont {Cora},
  \citenamefont {Fogarassy}, \citenamefont {Fornari}, \citenamefont {Bosi},
  \citenamefont {Re{\v c}nik},\ and\ \citenamefont
  {P{\'e}cz}}]{cora2020insitu}%
  \BibitemOpen
  \bibfield  {author} {\bibinfo {author} {\bibfnamefont {I.}~\bibnamefont
  {Cora}}, \bibinfo {author} {\bibfnamefont {Z.}~\bibnamefont {Fogarassy}},
  \bibinfo {author} {\bibfnamefont {R.}~\bibnamefont {Fornari}}, \bibinfo
  {author} {\bibfnamefont {M.}~\bibnamefont {Bosi}}, \bibinfo {author}
  {\bibfnamefont {A.}~\bibnamefont {Re{\v c}nik}},\ and\ \bibinfo {author}
  {\bibfnamefont {B.}~\bibnamefont {P{\'e}cz}},\ }\href
  {https://doi.org/10.1016/j.actamat.2019.11.019} {\bibfield  {journal}
  {\bibinfo  {journal} {Acta Mater.}\ }\textbf {\bibinfo {volume} {183}},\
  \bibinfo {pages} {216} (\bibinfo {year} {2020})}\BibitemShut {NoStop}%
\bibitem [{\citenamefont {Tetelbaum}\ \emph {et~al.}(2021)\citenamefont
  {Tetelbaum}, \citenamefont {Nikolskaya}, \citenamefont {Korolev},
  \citenamefont {Mullagaliev}, \citenamefont {Belov}, \citenamefont {Trushin},
  \citenamefont {Dudin}, \citenamefont {Nezhdanov}, \citenamefont {Mashin},
  \citenamefont {Mikhaylov}, \citenamefont {Pechnikov}, \citenamefont
  {Scheglov}, \citenamefont {Nikolaev},\ and\ \citenamefont
  {Gogova}}]{tetelbaum2021ionbeam}%
  \BibitemOpen
  \bibfield  {author} {\bibinfo {author} {\bibfnamefont {D.}~\bibnamefont
  {Tetelbaum}}, \bibinfo {author} {\bibfnamefont {A.}~\bibnamefont
  {Nikolskaya}}, \bibinfo {author} {\bibfnamefont {D.}~\bibnamefont {Korolev}},
  \bibinfo {author} {\bibfnamefont {T.}~\bibnamefont {Mullagaliev}}, \bibinfo
  {author} {\bibfnamefont {A.}~\bibnamefont {Belov}}, \bibinfo {author}
  {\bibfnamefont {V.}~\bibnamefont {Trushin}}, \bibinfo {author} {\bibfnamefont
  {Y.}~\bibnamefont {Dudin}}, \bibinfo {author} {\bibfnamefont
  {A.}~\bibnamefont {Nezhdanov}}, \bibinfo {author} {\bibfnamefont
  {A.}~\bibnamefont {Mashin}}, \bibinfo {author} {\bibfnamefont
  {A.}~\bibnamefont {Mikhaylov}}, \bibinfo {author} {\bibfnamefont
  {A.}~\bibnamefont {Pechnikov}}, \bibinfo {author} {\bibfnamefont
  {M.}~\bibnamefont {Scheglov}}, \bibinfo {author} {\bibfnamefont
  {V.}~\bibnamefont {Nikolaev}},\ and\ \bibinfo {author} {\bibfnamefont
  {D.}~\bibnamefont {Gogova}},\ }\href
  {https://doi.org/https://doi.org/10.1016/j.matlet.2021.130346} {\bibfield
  {journal} {\bibinfo  {journal} {Mater. Lett.}\ }\textbf {\bibinfo {volume}
  {302}},\ \bibinfo {pages} {130346} (\bibinfo {year} {2021})}\BibitemShut
  {NoStop}%
\bibitem [{\citenamefont {Azarov}\ \emph {et~al.}(2022)\citenamefont {Azarov},
  \citenamefont {Bazioti}, \citenamefont {Venkatachalapathy}, \citenamefont
  {Vajeeston}, \citenamefont {Monakhov},\ and\ \citenamefont
  {Kuznetsov}}]{azarov2022disorder}%
  \BibitemOpen
  \bibfield  {author} {\bibinfo {author} {\bibfnamefont {A.}~\bibnamefont
  {Azarov}}, \bibinfo {author} {\bibfnamefont {C.}~\bibnamefont {Bazioti}},
  \bibinfo {author} {\bibfnamefont {V.}~\bibnamefont {Venkatachalapathy}},
  \bibinfo {author} {\bibfnamefont {P.}~\bibnamefont {Vajeeston}}, \bibinfo
  {author} {\bibfnamefont {E.}~\bibnamefont {Monakhov}},\ and\ \bibinfo
  {author} {\bibfnamefont {A.}~\bibnamefont {Kuznetsov}},\ }\href
  {https://doi.org/10.1103/PhysRevLett.128.015704} {\bibfield  {journal}
  {\bibinfo  {journal} {Phys. Rev. Lett.}\ }\textbf {\bibinfo {volume} {128}},\
  \bibinfo {pages} {015704} (\bibinfo {year} {2022})}\BibitemShut {NoStop}%
\bibitem [{\citenamefont {Zhu}\ \emph {et~al.}(2016)\citenamefont {Zhu},
  \citenamefont {Guan},\ and\ \citenamefont {Liu}}]{zhu2016mechanism}%
  \BibitemOpen
  \bibfield  {author} {\bibinfo {author} {\bibfnamefont {S.-C.}\ \bibnamefont
  {Zhu}}, \bibinfo {author} {\bibfnamefont {S.-H.}\ \bibnamefont {Guan}},\ and\
  \bibinfo {author} {\bibfnamefont {Z.-P.}\ \bibnamefont {Liu}},\ }\href
  {https://doi.org/10.1039/C6CP03673B} {\bibfield  {journal} {\bibinfo
  {journal} {Phys. Chem. Chem. Phys.}\ }\textbf {\bibinfo {volume} {18}},\
  \bibinfo {pages} {18563} (\bibinfo {year} {2016})}\BibitemShut {NoStop}%
\bibitem [{\citenamefont {Schubert}\ \emph {et~al.}(2016)\citenamefont
  {Schubert}, \citenamefont {Korlacki}, \citenamefont {Knight}, \citenamefont
  {Hofmann}, \citenamefont {Sch{\"o}che}, \citenamefont {Darakchieva},
  \citenamefont {Janz{\'e}n}, \citenamefont {Monemar}, \citenamefont {Gogova},
  \citenamefont {Thieu}, \citenamefont {Togashi}, \citenamefont {Murakami},
  \citenamefont {Kumagai}, \citenamefont {Goto}, \citenamefont {Kuramata},
  \citenamefont {Yamakoshi},\ and\ \citenamefont
  {Higashiwaki}}]{schubert2016anisotropy}%
  \BibitemOpen
  \bibfield  {author} {\bibinfo {author} {\bibfnamefont {M.}~\bibnamefont
  {Schubert}}, \bibinfo {author} {\bibfnamefont {R.}~\bibnamefont {Korlacki}},
  \bibinfo {author} {\bibfnamefont {S.}~\bibnamefont {Knight}}, \bibinfo
  {author} {\bibfnamefont {T.}~\bibnamefont {Hofmann}}, \bibinfo {author}
  {\bibfnamefont {S.}~\bibnamefont {Sch{\"o}che}}, \bibinfo {author}
  {\bibfnamefont {V.}~\bibnamefont {Darakchieva}}, \bibinfo {author}
  {\bibfnamefont {E.}~\bibnamefont {Janz{\'e}n}}, \bibinfo {author}
  {\bibfnamefont {B.}~\bibnamefont {Monemar}}, \bibinfo {author} {\bibfnamefont
  {D.}~\bibnamefont {Gogova}}, \bibinfo {author} {\bibfnamefont {Q.-T.}\
  \bibnamefont {Thieu}}, \bibinfo {author} {\bibfnamefont {R.}~\bibnamefont
  {Togashi}}, \bibinfo {author} {\bibfnamefont {H.}~\bibnamefont {Murakami}},
  \bibinfo {author} {\bibfnamefont {Y.}~\bibnamefont {Kumagai}}, \bibinfo
  {author} {\bibfnamefont {K.}~\bibnamefont {Goto}}, \bibinfo {author}
  {\bibfnamefont {A.}~\bibnamefont {Kuramata}}, \bibinfo {author}
  {\bibfnamefont {S.}~\bibnamefont {Yamakoshi}},\ and\ \bibinfo {author}
  {\bibfnamefont {M.}~\bibnamefont {Higashiwaki}},\ }\href
  {https://doi.org/10.1103/PhysRevB.93.125209} {\bibfield  {journal} {\bibinfo
  {journal} {Phys. Rev. B}\ }\textbf {\bibinfo {volume} {93}},\ \bibinfo
  {pages} {125209} (\bibinfo {year} {2016})}\BibitemShut {NoStop}%
\bibitem [{\citenamefont {Furthm{\"u}ller}\ and\ \citenamefont
  {Bechstedt}(2016)}]{furthm2016quasi}%
  \BibitemOpen
  \bibfield  {author} {\bibinfo {author} {\bibfnamefont {J.}~\bibnamefont
  {Furthm{\"u}ller}}\ and\ \bibinfo {author} {\bibfnamefont {F.}~\bibnamefont
  {Bechstedt}},\ }\href {https://doi.org/10.1103/PhysRevB.93.115204} {\bibfield
   {journal} {\bibinfo  {journal} {Phys. Rev. B}\ }\textbf {\bibinfo {volume}
  {93}},\ \bibinfo {pages} {115204} (\bibinfo {year} {2016})}\BibitemShut
  {NoStop}%
\bibitem [{\citenamefont {Ponc{\'e}}\ and\ \citenamefont
  {Giustino}(2020)}]{ponce2020structural}%
  \BibitemOpen
  \bibfield  {author} {\bibinfo {author} {\bibfnamefont {S.}~\bibnamefont
  {Ponc{\'e}}}\ and\ \bibinfo {author} {\bibfnamefont {F.}~\bibnamefont
  {Giustino}},\ }\href {https://doi.org/10.1103/PhysRevResearch.2.033102}
  {\bibfield  {journal} {\bibinfo  {journal} {Phys. Rev. Res.}\ }\textbf
  {\bibinfo {volume} {2}},\ \bibinfo {pages} {033102} (\bibinfo {year}
  {2020})}\BibitemShut {NoStop}%
\bibitem [{\citenamefont {Varley}\ \emph {et~al.}(2016)\citenamefont {Varley},
  \citenamefont {Weber}, \citenamefont {Janotti},\ and\ \citenamefont {Van~de
  Walle}}]{varley2016oxygenerr}%
  \BibitemOpen
  \bibfield  {author} {\bibinfo {author} {\bibfnamefont {J.~B.}\ \bibnamefont
  {Varley}}, \bibinfo {author} {\bibfnamefont {J.~R.}\ \bibnamefont {Weber}},
  \bibinfo {author} {\bibfnamefont {A.}~\bibnamefont {Janotti}},\ and\ \bibinfo
  {author} {\bibfnamefont {C.~G.}\ \bibnamefont {Van~de Walle}},\ }\href
  {https://doi.org/10.1063/1.4940444} {\bibfield  {journal} {\bibinfo
  {journal} {Appl. Phys. Lett.}\ }\textbf {\bibinfo {volume} {108}},\ \bibinfo
  {pages} {039901} (\bibinfo {year} {2016})}\BibitemShut {NoStop}%
\bibitem [{\citenamefont {Peelaers}\ \emph {et~al.}(2019)\citenamefont
  {Peelaers}, \citenamefont {Lyons}, \citenamefont {Varley},\ and\
  \citenamefont {Van~de Walle}}]{peelaers2019deep}%
  \BibitemOpen
  \bibfield  {author} {\bibinfo {author} {\bibfnamefont {H.}~\bibnamefont
  {Peelaers}}, \bibinfo {author} {\bibfnamefont {J.~L.}\ \bibnamefont {Lyons}},
  \bibinfo {author} {\bibfnamefont {J.~B.}\ \bibnamefont {Varley}},\ and\
  \bibinfo {author} {\bibfnamefont {C.~G.}\ \bibnamefont {Van~de Walle}},\
  }\href {https://doi.org/10.1063/1.5063807} {\bibfield  {journal} {\bibinfo
  {journal} {APL Mater.}\ }\textbf {\bibinfo {volume} {7}},\ \bibinfo {pages}
  {022519} (\bibinfo {year} {2019})}\BibitemShut {NoStop}%
\bibitem [{\citenamefont {Peelaers}\ and\ \citenamefont {Van~de
  Walle}(2017)}]{peelaers2017lack}%
  \BibitemOpen
  \bibfield  {author} {\bibinfo {author} {\bibfnamefont {H.}~\bibnamefont
  {Peelaers}}\ and\ \bibinfo {author} {\bibfnamefont {C.~G.}\ \bibnamefont
  {Van~de Walle}},\ }\href {https://doi.org/10.1103/PhysRevB.96.081409}
  {\bibfield  {journal} {\bibinfo  {journal} {Phys. Rev. B}\ }\textbf {\bibinfo
  {volume} {96}},\ \bibinfo {pages} {081409(R)} (\bibinfo {year}
  {2017})}\BibitemShut {NoStop}%
\bibitem [{\citenamefont {Mu}\ \emph {et~al.}(2020)\citenamefont {Mu},
  \citenamefont {Wang}, \citenamefont {Peelaers},\ and\ \citenamefont {Van~de
  Walle}}]{mu2020first}%
  \BibitemOpen
  \bibfield  {author} {\bibinfo {author} {\bibfnamefont {S.}~\bibnamefont
  {Mu}}, \bibinfo {author} {\bibfnamefont {M.}~\bibnamefont {Wang}}, \bibinfo
  {author} {\bibfnamefont {H.}~\bibnamefont {Peelaers}},\ and\ \bibinfo
  {author} {\bibfnamefont {C.~G.}\ \bibnamefont {Van~de Walle}},\ }\href
  {https://doi.org/10.1063/5.0019915} {\bibfield  {journal} {\bibinfo
  {journal} {APL Mater.}\ }\textbf {\bibinfo {volume} {8}},\ \bibinfo {pages}
  {091105} (\bibinfo {year} {2020})}\BibitemShut {NoStop}%
\bibitem [{\citenamefont {Deringer}\ \emph {et~al.}(2021)\citenamefont
  {Deringer}, \citenamefont {Bart{\'o}k}, \citenamefont {Bernstein},
  \citenamefont {Wilkins}, \citenamefont {Ceriotti},\ and\ \citenamefont
  {Cs{\'a}nyi}}]{deringer2021gaussian}%
  \BibitemOpen
  \bibfield  {author} {\bibinfo {author} {\bibfnamefont {V.~L.}\ \bibnamefont
  {Deringer}}, \bibinfo {author} {\bibfnamefont {A.~P.}\ \bibnamefont
  {Bart{\'o}k}}, \bibinfo {author} {\bibfnamefont {N.}~\bibnamefont
  {Bernstein}}, \bibinfo {author} {\bibfnamefont {D.~M.}\ \bibnamefont
  {Wilkins}}, \bibinfo {author} {\bibfnamefont {M.}~\bibnamefont {Ceriotti}},\
  and\ \bibinfo {author} {\bibfnamefont {G.}~\bibnamefont {Cs{\'a}nyi}},\
  }\href {https://doi.org/10.1021/acs.chemrev.1c00022} {\bibfield  {journal}
  {\bibinfo  {journal} {Chem. Rev.}\ }\textbf {\bibinfo {volume} {121}},\
  \bibinfo {pages} {10073} (\bibinfo {year} {2021})}\BibitemShut {NoStop}%
\bibitem [{\citenamefont {Friederich}\ \emph {et~al.}(2021)\citenamefont
  {Friederich}, \citenamefont {H{\"a}se}, \citenamefont {Proppe},\ and\
  \citenamefont {Aspuru-Guzik}}]{friederich2021machine}%
  \BibitemOpen
  \bibfield  {author} {\bibinfo {author} {\bibfnamefont {P.}~\bibnamefont
  {Friederich}}, \bibinfo {author} {\bibfnamefont {F.}~\bibnamefont
  {H{\"a}se}}, \bibinfo {author} {\bibfnamefont {J.}~\bibnamefont {Proppe}},\
  and\ \bibinfo {author} {\bibfnamefont {A.}~\bibnamefont {Aspuru-Guzik}},\
  }\href {https://doi.org/10.1038/s41563-020-0777-6} {\bibfield  {journal}
  {\bibinfo  {journal} {Nat. Mater.}\ }\textbf {\bibinfo {volume} {20}},\
  \bibinfo {pages} {750} (\bibinfo {year} {2021})}\BibitemShut {NoStop}%
\bibitem [{\citenamefont {Bart{\'o}k}\ \emph {et~al.}(2010)\citenamefont
  {Bart{\'o}k}, \citenamefont {Payne}, \citenamefont {Kondor},\ and\
  \citenamefont {Cs{\'a}nyi}}]{gap2010}%
  \BibitemOpen
  \bibfield  {author} {\bibinfo {author} {\bibfnamefont {A.~P.}\ \bibnamefont
  {Bart{\'o}k}}, \bibinfo {author} {\bibfnamefont {M.~C.}\ \bibnamefont
  {Payne}}, \bibinfo {author} {\bibfnamefont {R.}~\bibnamefont {Kondor}},\ and\
  \bibinfo {author} {\bibfnamefont {G.}~\bibnamefont {Cs{\'a}nyi}},\ }\href
  {https://doi.org/10.1103/PhysRevLett.104.136403} {\bibfield  {journal}
  {\bibinfo  {journal} {Phys. Rev. Lett.}\ }\textbf {\bibinfo {volume} {104}},\
  \bibinfo {pages} {136403} (\bibinfo {year} {2010})}\BibitemShut {NoStop}%
\bibitem [{\citenamefont {Bart{\'o}k}\ \emph {et~al.}(2018)\citenamefont
  {Bart{\'o}k}, \citenamefont {Kermode}, \citenamefont {Bernstein},\ and\
  \citenamefont {Cs{\'a}nyi}}]{bartok2018machine}%
  \BibitemOpen
  \bibfield  {author} {\bibinfo {author} {\bibfnamefont {A.~P.}\ \bibnamefont
  {Bart{\'o}k}}, \bibinfo {author} {\bibfnamefont {J.}~\bibnamefont {Kermode}},
  \bibinfo {author} {\bibfnamefont {N.}~\bibnamefont {Bernstein}},\ and\
  \bibinfo {author} {\bibfnamefont {G.}~\bibnamefont {Cs{\'a}nyi}},\ }\href
  {https://doi.org/10.1103/PhysRevX.8.041048} {\bibfield  {journal} {\bibinfo
  {journal} {Phys. Rev. X}\ }\textbf {\bibinfo {volume} {8}},\ \bibinfo {pages}
  {041048} (\bibinfo {year} {2018})}\BibitemShut {NoStop}%
\bibitem [{\citenamefont {Rowe}\ \emph {et~al.}(2020)\citenamefont {Rowe},
  \citenamefont {Deringer}, \citenamefont {Gasparotto}, \citenamefont
  {Cs{\'a}nyi},\ and\ \citenamefont {Michaelides}}]{rowe2020carbon}%
  \BibitemOpen
  \bibfield  {author} {\bibinfo {author} {\bibfnamefont {P.}~\bibnamefont
  {Rowe}}, \bibinfo {author} {\bibfnamefont {V.~L.}\ \bibnamefont {Deringer}},
  \bibinfo {author} {\bibfnamefont {P.}~\bibnamefont {Gasparotto}}, \bibinfo
  {author} {\bibfnamefont {G.}~\bibnamefont {Cs{\'a}nyi}},\ and\ \bibinfo
  {author} {\bibfnamefont {A.}~\bibnamefont {Michaelides}},\ }\href
  {https://doi.org/10.1063/5.0005084} {\bibfield  {journal} {\bibinfo
  {journal} {J. Chem. Phys.}\ }\textbf {\bibinfo {volume} {153}},\ \bibinfo
  {pages} {034702} (\bibinfo {year} {2020})}\BibitemShut {NoStop}%
\bibitem [{\citenamefont {Deringer}\ \emph {et~al.}(2020)\citenamefont
  {Deringer}, \citenamefont {Caro},\ and\ \citenamefont
  {Cs{\'a}nyi}}]{deringer2020a}%
  \BibitemOpen
  \bibfield  {author} {\bibinfo {author} {\bibfnamefont {V.~L.}\ \bibnamefont
  {Deringer}}, \bibinfo {author} {\bibfnamefont {M.~A.}\ \bibnamefont {Caro}},\
  and\ \bibinfo {author} {\bibfnamefont {G.}~\bibnamefont {Cs{\'a}nyi}},\
  }\href {https://doi.org/10.1038/s41467-020-19168-z} {\bibfield  {journal}
  {\bibinfo  {journal} {Nat. Commun.}\ }\textbf {\bibinfo {volume} {11}},\
  \bibinfo {pages} {5461} (\bibinfo {year} {2020})}\BibitemShut {NoStop}%
\bibitem [{\citenamefont {Sivaraman}\ \emph {et~al.}(2020)\citenamefont
  {Sivaraman}, \citenamefont {Krishnamoorthy}, \citenamefont {Baur},
  \citenamefont {Holm}, \citenamefont {Stan}, \citenamefont {Cs{\'a}nyi},
  \citenamefont {Benmore},\ and\ \citenamefont
  {V{\'a}zquez-Mayagoitia}}]{sivaraman2020machine}%
  \BibitemOpen
  \bibfield  {author} {\bibinfo {author} {\bibfnamefont {G.}~\bibnamefont
  {Sivaraman}}, \bibinfo {author} {\bibfnamefont {A.~N.}\ \bibnamefont
  {Krishnamoorthy}}, \bibinfo {author} {\bibfnamefont {M.}~\bibnamefont
  {Baur}}, \bibinfo {author} {\bibfnamefont {C.}~\bibnamefont {Holm}}, \bibinfo
  {author} {\bibfnamefont {M.}~\bibnamefont {Stan}}, \bibinfo {author}
  {\bibfnamefont {G.}~\bibnamefont {Cs{\'a}nyi}}, \bibinfo {author}
  {\bibfnamefont {C.}~\bibnamefont {Benmore}},\ and\ \bibinfo {author}
  {\bibfnamefont {{\'A}.}~\bibnamefont {V{\'a}zquez-Mayagoitia}},\ }\href
  {https://doi.org/10.1038/s41524-020-00367-7} {\bibfield  {journal} {\bibinfo
  {journal} {npj Comput. Mater.}\ }\textbf {\bibinfo {volume} {6}},\ \bibinfo
  {pages} {104} (\bibinfo {year} {2020})}\BibitemShut {NoStop}%
\bibitem [{\citenamefont {Erhard}\ \emph {et~al.}(2022)\citenamefont {Erhard},
  \citenamefont {Rohrer}, \citenamefont {Albe},\ and\ \citenamefont
  {Deringer}}]{erhard2022machine}%
  \BibitemOpen
  \bibfield  {author} {\bibinfo {author} {\bibfnamefont {L.~C.}\ \bibnamefont
  {Erhard}}, \bibinfo {author} {\bibfnamefont {J.}~\bibnamefont {Rohrer}},
  \bibinfo {author} {\bibfnamefont {K.}~\bibnamefont {Albe}},\ and\ \bibinfo
  {author} {\bibfnamefont {V.~L.}\ \bibnamefont {Deringer}},\ }\href
  {https://doi.org/10.1038/s41524-022-00768-w} {\bibfield  {journal} {\bibinfo
  {journal} {npj Comput. Mater.}\ }\textbf {\bibinfo {volume} {8}},\ \bibinfo
  {pages} {90} (\bibinfo {year} {2022})}\BibitemShut {NoStop}%
\bibitem [{\citenamefont {Liu}\ \emph {et~al.}(2020)\citenamefont {Liu},
  \citenamefont {Yang}, \citenamefont {Xin}, \citenamefont {Liu}, \citenamefont
  {Cs{\'a}nyi},\ and\ \citenamefont {Cao}}]{liu2020machine}%
  \BibitemOpen
  \bibfield  {author} {\bibinfo {author} {\bibfnamefont {Y.-B.}\ \bibnamefont
  {Liu}}, \bibinfo {author} {\bibfnamefont {J.-Y.}\ \bibnamefont {Yang}},
  \bibinfo {author} {\bibfnamefont {G.-M.}\ \bibnamefont {Xin}}, \bibinfo
  {author} {\bibfnamefont {L.-H.}\ \bibnamefont {Liu}}, \bibinfo {author}
  {\bibfnamefont {G.}~\bibnamefont {Cs{\'a}nyi}},\ and\ \bibinfo {author}
  {\bibfnamefont {B.-Y.}\ \bibnamefont {Cao}},\ }\href
  {https://doi.org/10.1063/5.0027643} {\bibfield  {journal} {\bibinfo
  {journal} {J. Chem. Phys.}\ }\textbf {\bibinfo {volume} {153}},\ \bibinfo
  {pages} {144501} (\bibinfo {year} {2020})}\BibitemShut {NoStop}%
\bibitem [{\citenamefont {Li}\ \emph {et~al.}(2020)\citenamefont {Li},
  \citenamefont {Liu}, \citenamefont {Rohskopf}, \citenamefont {Gordiz},
  \citenamefont {Henry}, \citenamefont {Lee},\ and\ \citenamefont
  {Luo}}]{li2020a}%
  \BibitemOpen
  \bibfield  {author} {\bibinfo {author} {\bibfnamefont {R.}~\bibnamefont
  {Li}}, \bibinfo {author} {\bibfnamefont {Z.}~\bibnamefont {Liu}}, \bibinfo
  {author} {\bibfnamefont {A.}~\bibnamefont {Rohskopf}}, \bibinfo {author}
  {\bibfnamefont {K.}~\bibnamefont {Gordiz}}, \bibinfo {author} {\bibfnamefont
  {A.}~\bibnamefont {Henry}}, \bibinfo {author} {\bibfnamefont
  {E.}~\bibnamefont {Lee}},\ and\ \bibinfo {author} {\bibfnamefont
  {T.}~\bibnamefont {Luo}},\ }\href {https://doi.org/10.1063/5.0025051}
  {\bibfield  {journal} {\bibinfo  {journal} {Appl. Phys. Lett.}\ }\textbf
  {\bibinfo {volume} {117}},\ \bibinfo {pages} {152102} (\bibinfo {year}
  {2020})}\BibitemShut {NoStop}%
\bibitem [{\citenamefont {Liu}\ \emph {et~al.}()\citenamefont {Liu},
  \citenamefont {Liang}, \citenamefont {Yang}, \citenamefont {Yang},
  \citenamefont {Yang}, \citenamefont {Song}, \citenamefont {Mei},
  \citenamefont {Cs{\'a}nyi},\ and\ \citenamefont {Cao}}]{liu2023unraveling}%
  \BibitemOpen
  \bibfield  {author} {\bibinfo {author} {\bibfnamefont {Y.}~\bibnamefont
  {Liu}}, \bibinfo {author} {\bibfnamefont {H.}~\bibnamefont {Liang}}, \bibinfo
  {author} {\bibfnamefont {L.}~\bibnamefont {Yang}}, \bibinfo {author}
  {\bibfnamefont {G.}~\bibnamefont {Yang}}, \bibinfo {author} {\bibfnamefont
  {H.}~\bibnamefont {Yang}}, \bibinfo {author} {\bibfnamefont {S.}~\bibnamefont
  {Song}}, \bibinfo {author} {\bibfnamefont {Z.}~\bibnamefont {Mei}}, \bibinfo
  {author} {\bibfnamefont {G.}~\bibnamefont {Cs{\'a}nyi}},\ and\ \bibinfo
  {author} {\bibfnamefont {B.}~\bibnamefont {Cao}},\ }\href
  {https://doi.org/10.1002/adma.202210873} {\bibfield  {journal} {\bibinfo
  {journal} {Adv. Mater.}\ }\textbf {\bibinfo {volume} {n/a}},\ \bibinfo
  {pages} {2210873}}\BibitemShut {NoStop}%
\bibitem [{\citenamefont {Zhao}\ \emph
  {et~al.}(2021{\natexlab{b}})\citenamefont {Zhao}, \citenamefont
  {Byggm{\"a}star}, \citenamefont {Zhang}, \citenamefont {Djurabekova},
  \citenamefont {Nordlund},\ and\ \citenamefont {Hua}}]{SFzhao2021phase}%
  \BibitemOpen
  \bibfield  {author} {\bibinfo {author} {\bibfnamefont {J.}~\bibnamefont
  {Zhao}}, \bibinfo {author} {\bibfnamefont {J.}~\bibnamefont
  {Byggm{\"a}star}}, \bibinfo {author} {\bibfnamefont {Z.}~\bibnamefont
  {Zhang}}, \bibinfo {author} {\bibfnamefont {F.}~\bibnamefont {Djurabekova}},
  \bibinfo {author} {\bibfnamefont {K.}~\bibnamefont {Nordlund}},\ and\
  \bibinfo {author} {\bibfnamefont {M.}~\bibnamefont {Hua}},\ }\href
  {https://doi.org/10.1103/PhysRevB.104.054107} {\bibfield  {journal} {\bibinfo
   {journal} {Phys. Rev. B}\ }\textbf {\bibinfo {volume} {104}},\ \bibinfo
  {pages} {054107} (\bibinfo {year} {2021}{\natexlab{b}})}\BibitemShut
  {NoStop}%
\bibitem [{\citenamefont {Wang}\ \emph {et~al.}(2020)\citenamefont {Wang},
  \citenamefont {Faizan}, \citenamefont {Na}, \citenamefont {He}, \citenamefont
  {Fu},\ and\ \citenamefont {Zhang}}]{wang2020discovery}%
  \BibitemOpen
  \bibfield  {author} {\bibinfo {author} {\bibfnamefont {X.}~\bibnamefont
  {Wang}}, \bibinfo {author} {\bibfnamefont {M.}~\bibnamefont {Faizan}},
  \bibinfo {author} {\bibfnamefont {G.}~\bibnamefont {Na}}, \bibinfo {author}
  {\bibfnamefont {X.}~\bibnamefont {He}}, \bibinfo {author} {\bibfnamefont
  {Y.}~\bibnamefont {Fu}},\ and\ \bibinfo {author} {\bibfnamefont
  {L.}~\bibnamefont {Zhang}},\ }\href
  {https://doi.org/https://doi.org/10.1002/aelm.202000119} {\bibfield
  {journal} {\bibinfo  {journal} {Adv. Electron. Mater.}\ }\textbf {\bibinfo
  {volume} {6}},\ \bibinfo {pages} {2000119} (\bibinfo {year}
  {2020})}\BibitemShut {NoStop}%
\bibitem [{\citenamefont {Swallow}\ \emph {et~al.}(2021)\citenamefont
  {Swallow}, \citenamefont {Palgrave}, \citenamefont {Murgatroyd},
  \citenamefont {Regoutz}, \citenamefont {Lorenz}, \citenamefont {Hassa},
  \citenamefont {Grundmann}, \citenamefont {von Wenckstern}, \citenamefont
  {Varley},\ and\ \citenamefont {Veal}}]{swallow2021indium}%
  \BibitemOpen
  \bibfield  {author} {\bibinfo {author} {\bibfnamefont {J.~E.~N.}\
  \bibnamefont {Swallow}}, \bibinfo {author} {\bibfnamefont {R.~G.}\
  \bibnamefont {Palgrave}}, \bibinfo {author} {\bibfnamefont {P.~A.~E.}\
  \bibnamefont {Murgatroyd}}, \bibinfo {author} {\bibfnamefont
  {A.}~\bibnamefont {Regoutz}}, \bibinfo {author} {\bibfnamefont
  {M.}~\bibnamefont {Lorenz}}, \bibinfo {author} {\bibfnamefont
  {A.}~\bibnamefont {Hassa}}, \bibinfo {author} {\bibfnamefont
  {M.}~\bibnamefont {Grundmann}}, \bibinfo {author} {\bibfnamefont
  {H.}~\bibnamefont {von Wenckstern}}, \bibinfo {author} {\bibfnamefont
  {J.~B.}\ \bibnamefont {Varley}},\ and\ \bibinfo {author} {\bibfnamefont
  {T.~D.}\ \bibnamefont {Veal}},\ }\href
  {https://doi.org/10.1021/acsami.0c16021} {\bibfield  {journal} {\bibinfo
  {journal} {ACS Appl. Mater. Interfaces}\ }\textbf {\bibinfo {volume} {13}},\
  \bibinfo {pages} {2807} (\bibinfo {year} {2021})}\BibitemShut {NoStop}%
\bibitem [{\citenamefont {Jain}\ \emph {et~al.}(2013)\citenamefont {Jain},
  \citenamefont {Ong}, \citenamefont {Hautier}, \citenamefont {Chen},
  \citenamefont {Richards}, \citenamefont {Dacek}, \citenamefont {Cholia},
  \citenamefont {Gunter}, \citenamefont {Skinner}, \citenamefont {Ceder},\ and\
  \citenamefont {Persson}}]{jain2013commentary}%
  \BibitemOpen
  \bibfield  {author} {\bibinfo {author} {\bibfnamefont {A.}~\bibnamefont
  {Jain}}, \bibinfo {author} {\bibfnamefont {S.~P.}\ \bibnamefont {Ong}},
  \bibinfo {author} {\bibfnamefont {G.}~\bibnamefont {Hautier}}, \bibinfo
  {author} {\bibfnamefont {W.}~\bibnamefont {Chen}}, \bibinfo {author}
  {\bibfnamefont {W.~D.}\ \bibnamefont {Richards}}, \bibinfo {author}
  {\bibfnamefont {S.}~\bibnamefont {Dacek}}, \bibinfo {author} {\bibfnamefont
  {S.}~\bibnamefont {Cholia}}, \bibinfo {author} {\bibfnamefont
  {D.}~\bibnamefont {Gunter}}, \bibinfo {author} {\bibfnamefont
  {D.}~\bibnamefont {Skinner}}, \bibinfo {author} {\bibfnamefont
  {G.}~\bibnamefont {Ceder}},\ and\ \bibinfo {author} {\bibfnamefont {K.~A.}\
  \bibnamefont {Persson}},\ }\href {https://doi.org/10.1063/1.4812323}
  {\bibfield  {journal} {\bibinfo  {journal} {APL Mater.}\ }\textbf {\bibinfo
  {volume} {1}},\ \bibinfo {pages} {011002} (\bibinfo {year}
  {2013})}\BibitemShut {NoStop}%
\bibitem [{\citenamefont {Nordlund}\ \emph {et~al.}(1997)\citenamefont
  {Nordlund}, \citenamefont {Runeberg},\ and\ \citenamefont
  {Sundholm}}]{nordlund1997repulsive}%
  \BibitemOpen
  \bibfield  {author} {\bibinfo {author} {\bibfnamefont {K.}~\bibnamefont
  {Nordlund}}, \bibinfo {author} {\bibfnamefont {N.}~\bibnamefont {Runeberg}},\
  and\ \bibinfo {author} {\bibfnamefont {D.}~\bibnamefont {Sundholm}},\ }\href
  {https://doi.org/10.1016/S0168-583X(97)00447-3} {\bibfield  {journal}
  {\bibinfo  {journal} {Nucl. Instrum. Methods Phys. Res., Sect. B}\ }\textbf
  {\bibinfo {volume} {132}},\ \bibinfo {pages} {45} (\bibinfo {year}
  {1997})}\BibitemShut {NoStop}%
\bibitem [{\citenamefont {Byggm{\"a}star}\ \emph {et~al.}(2019)\citenamefont
  {Byggm{\"a}star}, \citenamefont {Hamedani}, \citenamefont {Nordlund},\ and\
  \citenamefont {Djurabekova}}]{byggmastar2019machine}%
  \BibitemOpen
  \bibfield  {author} {\bibinfo {author} {\bibfnamefont {J.}~\bibnamefont
  {Byggm{\"a}star}}, \bibinfo {author} {\bibfnamefont {A.}~\bibnamefont
  {Hamedani}}, \bibinfo {author} {\bibfnamefont {K.}~\bibnamefont {Nordlund}},\
  and\ \bibinfo {author} {\bibfnamefont {F.}~\bibnamefont {Djurabekova}},\
  }\href {https://doi.org/10.1103/PhysRevB.100.144105} {\bibfield  {journal}
  {\bibinfo  {journal} {Phys. Rev. B}\ }\textbf {\bibinfo {volume} {100}},\
  \bibinfo {pages} {144105} (\bibinfo {year} {2019})}\BibitemShut {NoStop}%
\bibitem [{\citenamefont {Bart{\'o}k}\ \emph {et~al.}(2013)\citenamefont
  {Bart{\'o}k}, \citenamefont {Kondor},\ and\ \citenamefont
  {Cs{\'a}nyi}}]{soap2013}%
  \BibitemOpen
  \bibfield  {author} {\bibinfo {author} {\bibfnamefont {A.~P.}\ \bibnamefont
  {Bart{\'o}k}}, \bibinfo {author} {\bibfnamefont {R.}~\bibnamefont {Kondor}},\
  and\ \bibinfo {author} {\bibfnamefont {G.}~\bibnamefont {Cs{\'a}nyi}},\
  }\href {https://doi.org/10.1103/PhysRevB.87.184115} {\bibfield  {journal}
  {\bibinfo  {journal} {Phys. Rev. B}\ }\textbf {\bibinfo {volume} {87}},\
  \bibinfo {pages} {184115} (\bibinfo {year} {2013})}\BibitemShut {NoStop}%
\bibitem [{\citenamefont {Pozdnyakov}\ \emph {et~al.}(2020)\citenamefont
  {Pozdnyakov}, \citenamefont {Willatt}, \citenamefont {Bart{\'o}k},
  \citenamefont {Ortner}, \citenamefont {Cs{\'a}nyi},\ and\ \citenamefont
  {Ceriotti}}]{twobody2020}%
  \BibitemOpen
  \bibfield  {author} {\bibinfo {author} {\bibfnamefont {S.~N.}\ \bibnamefont
  {Pozdnyakov}}, \bibinfo {author} {\bibfnamefont {M.~J.}\ \bibnamefont
  {Willatt}}, \bibinfo {author} {\bibfnamefont {A.~P.}\ \bibnamefont
  {Bart{\'o}k}}, \bibinfo {author} {\bibfnamefont {C.}~\bibnamefont {Ortner}},
  \bibinfo {author} {\bibfnamefont {G.}~\bibnamefont {Cs{\'a}nyi}},\ and\
  \bibinfo {author} {\bibfnamefont {M.}~\bibnamefont {Ceriotti}},\ }\href
  {https://doi.org/10.1103/PhysRevLett.125.166001} {\bibfield  {journal}
  {\bibinfo  {journal} {Phys. Rev. Lett.}\ }\textbf {\bibinfo {volume} {125}},\
  \bibinfo {pages} {166001} (\bibinfo {year} {2020})}\BibitemShut {NoStop}%
\bibitem [{\citenamefont {Byggm{\"a}star}\ \emph
  {et~al.}(2022{\natexlab{a}})\citenamefont {Byggm{\"a}star}, \citenamefont
  {Nordlund},\ and\ \citenamefont {Djurabekova}}]{byggmastar2022simple}%
  \BibitemOpen
  \bibfield  {author} {\bibinfo {author} {\bibfnamefont {J.}~\bibnamefont
  {Byggm{\"a}star}}, \bibinfo {author} {\bibfnamefont {K.}~\bibnamefont
  {Nordlund}},\ and\ \bibinfo {author} {\bibfnamefont {F.}~\bibnamefont
  {Djurabekova}},\ }\href {https://doi.org/10.1103/PhysRevMaterials.6.083801}
  {\bibfield  {journal} {\bibinfo  {journal} {Phys. Rev. Mater.}\ }\textbf
  {\bibinfo {volume} {6}},\ \bibinfo {pages} {083801} (\bibinfo {year}
  {2022}{\natexlab{a}})}\BibitemShut {NoStop}%
\bibitem [{\citenamefont {Bart{\'o}k}\ and\ \citenamefont
  {Cs{\'a}nyi}(2015)}]{bartok2015gaussian}%
  \BibitemOpen
  \bibfield  {author} {\bibinfo {author} {\bibfnamefont {A.~P.}\ \bibnamefont
  {Bart{\'o}k}}\ and\ \bibinfo {author} {\bibfnamefont {G.}~\bibnamefont
  {Cs{\'a}nyi}},\ }\href {https://doi.org/10.1002/qua.24927} {\bibfield
  {journal} {\bibinfo  {journal} {Int. J. Quantum Chem.}\ }\textbf {\bibinfo
  {volume} {115}},\ \bibinfo {pages} {1051} (\bibinfo {year}
  {2015})}\BibitemShut {NoStop}%
\bibitem [{\citenamefont {{\AA}hman}\ \emph {et~al.}(1996)\citenamefont
  {{\AA}hman}, \citenamefont {Svensson},\ and\ \citenamefont
  {Albertsson}}]{ahman1996a}%
  \BibitemOpen
  \bibfield  {author} {\bibinfo {author} {\bibfnamefont {J.}~\bibnamefont
  {{\AA}hman}}, \bibinfo {author} {\bibfnamefont {G.}~\bibnamefont
  {Svensson}},\ and\ \bibinfo {author} {\bibfnamefont {J.}~\bibnamefont
  {Albertsson}},\ }\href {https://doi.org/10.1107/S0108270195016404} {\bibfield
   {journal} {\bibinfo  {journal} {Acta Crystallogr., Sect. C: Struct. Chem.}\
  }\textbf {\bibinfo {volume} {52}},\ \bibinfo {pages} {1336} (\bibinfo {year}
  {1996})}\BibitemShut {NoStop}%
\bibitem [{\citenamefont {Zhang}\ \emph {et~al.}(2021)\citenamefont {Zhang},
  \citenamefont {Xu}, \citenamefont {Zhao}, \citenamefont {Wang}, \citenamefont
  {Hao}, \citenamefont {Zhang}, \citenamefont {Cheng},\ and\ \citenamefont
  {Dong}}]{SFzhang2021temperature}%
  \BibitemOpen
  \bibfield  {author} {\bibinfo {author} {\bibfnamefont {K.}~\bibnamefont
  {Zhang}}, \bibinfo {author} {\bibfnamefont {Z.}~\bibnamefont {Xu}}, \bibinfo
  {author} {\bibfnamefont {J.}~\bibnamefont {Zhao}}, \bibinfo {author}
  {\bibfnamefont {H.}~\bibnamefont {Wang}}, \bibinfo {author} {\bibfnamefont
  {J.}~\bibnamefont {Hao}}, \bibinfo {author} {\bibfnamefont {S.}~\bibnamefont
  {Zhang}}, \bibinfo {author} {\bibfnamefont {H.}~\bibnamefont {Cheng}},\ and\
  \bibinfo {author} {\bibfnamefont {B.}~\bibnamefont {Dong}},\ }\href
  {https://doi.org/10.1016/j.jallcom.2021.160665} {\bibfield  {journal}
  {\bibinfo  {journal} {J. Alloys Compd.}\ ,\ \bibinfo {pages} {160665}}
  (\bibinfo {year} {2021})}\BibitemShut {NoStop}%
\bibitem [{\citenamefont {Marezio}\ and\ \citenamefont
  {Remeika}(1967)}]{marezio1967bond}%
  \BibitemOpen
  \bibfield  {author} {\bibinfo {author} {\bibfnamefont {M.}~\bibnamefont
  {Marezio}}\ and\ \bibinfo {author} {\bibfnamefont {J.~P.}\ \bibnamefont
  {Remeika}},\ }\href {https://doi.org/10.1063/1.1840945} {\bibfield  {journal}
  {\bibinfo  {journal} {J. Chem. Phys.}\ }\textbf {\bibinfo {volume} {46}},\
  \bibinfo {pages} {1862} (\bibinfo {year} {1967})}\BibitemShut {NoStop}%
\bibitem [{\citenamefont {Kato}\ \emph {et~al.}(2023)\citenamefont {Kato},
  \citenamefont {Nishinaka}, \citenamefont {Shimazoe}, \citenamefont
  {Kanegae},\ and\ \citenamefont {Yoshimoto}}]{kato2023demonstration}%
  \BibitemOpen
  \bibfield  {author} {\bibinfo {author} {\bibfnamefont {T.}~\bibnamefont
  {Kato}}, \bibinfo {author} {\bibfnamefont {H.}~\bibnamefont {Nishinaka}},
  \bibinfo {author} {\bibfnamefont {K.}~\bibnamefont {Shimazoe}}, \bibinfo
  {author} {\bibfnamefont {K.}~\bibnamefont {Kanegae}},\ and\ \bibinfo {author}
  {\bibfnamefont {M.}~\bibnamefont {Yoshimoto}},\ }\href
  {https://doi.org/10.1021/acsaelm.2c01750} {\bibfield  {journal} {\bibinfo
  {journal} {ACS Appl. Electron. Mater.}\ }\textbf {\bibinfo {volume} {5}},\
  \bibinfo {pages} {1715} (\bibinfo {year} {2023})}\BibitemShut {NoStop}%
\bibitem [{\citenamefont {Playford}\ \emph {et~al.}(2014)\citenamefont
  {Playford}, \citenamefont {Hannon}, \citenamefont {Tucker}, \citenamefont
  {Dawson}, \citenamefont {Ashbrook}, \citenamefont {Kastiban}, \citenamefont
  {Sloan},\ and\ \citenamefont {Walton}}]{playford2014characterization}%
  \BibitemOpen
  \bibfield  {author} {\bibinfo {author} {\bibfnamefont {H.~Y.}\ \bibnamefont
  {Playford}}, \bibinfo {author} {\bibfnamefont {A.~C.}\ \bibnamefont
  {Hannon}}, \bibinfo {author} {\bibfnamefont {M.~G.}\ \bibnamefont {Tucker}},
  \bibinfo {author} {\bibfnamefont {D.~M.}\ \bibnamefont {Dawson}}, \bibinfo
  {author} {\bibfnamefont {S.~E.}\ \bibnamefont {Ashbrook}}, \bibinfo {author}
  {\bibfnamefont {R.~J.}\ \bibnamefont {Kastiban}}, \bibinfo {author}
  {\bibfnamefont {J.}~\bibnamefont {Sloan}},\ and\ \bibinfo {author}
  {\bibfnamefont {R.~I.}\ \bibnamefont {Walton}},\ }\href
  {https://doi.org/10.1021/jp5033806} {\bibfield  {journal} {\bibinfo
  {journal} {J. Phys. Chem. C}\ }\textbf {\bibinfo {volume} {118}},\ \bibinfo
  {pages} {16188} (\bibinfo {year} {2014})}\BibitemShut {NoStop}%
\bibitem [{\citenamefont {Ratcliff}\ \emph {et~al.}(2022)\citenamefont
  {Ratcliff}, \citenamefont {Oshima}, \citenamefont {Nippert}, \citenamefont
  {Janzen}, \citenamefont {Kluth}, \citenamefont {Goldhahn}, \citenamefont
  {Feneberg}, \citenamefont {Mazzolini}, \citenamefont {Bierwagen},
  \citenamefont {Wouters}, \citenamefont {Nofal}, \citenamefont {Albrecht},
  \citenamefont {Swallow}, \citenamefont {Jones}, \citenamefont {Thakur},
  \citenamefont {Lee}, \citenamefont {Kalha}, \citenamefont {Schlueter},
  \citenamefont {Veal}, \citenamefont {Varley}, \citenamefont {Wagner},\ and\
  \citenamefont {Regoutz}}]{ratcliff2022tackling}%
  \BibitemOpen
  \bibfield  {author} {\bibinfo {author} {\bibfnamefont {L.~E.}\ \bibnamefont
  {Ratcliff}}, \bibinfo {author} {\bibfnamefont {T.}~\bibnamefont {Oshima}},
  \bibinfo {author} {\bibfnamefont {F.}~\bibnamefont {Nippert}}, \bibinfo
  {author} {\bibfnamefont {B.~M.}\ \bibnamefont {Janzen}}, \bibinfo {author}
  {\bibfnamefont {E.}~\bibnamefont {Kluth}}, \bibinfo {author} {\bibfnamefont
  {R.}~\bibnamefont {Goldhahn}}, \bibinfo {author} {\bibfnamefont
  {M.}~\bibnamefont {Feneberg}}, \bibinfo {author} {\bibfnamefont
  {P.}~\bibnamefont {Mazzolini}}, \bibinfo {author} {\bibfnamefont
  {O.}~\bibnamefont {Bierwagen}}, \bibinfo {author} {\bibfnamefont
  {C.}~\bibnamefont {Wouters}}, \bibinfo {author} {\bibfnamefont
  {M.}~\bibnamefont {Nofal}}, \bibinfo {author} {\bibfnamefont
  {M.}~\bibnamefont {Albrecht}}, \bibinfo {author} {\bibfnamefont {J.~E.~N.}\
  \bibnamefont {Swallow}}, \bibinfo {author} {\bibfnamefont {L.~A.~H.}\
  \bibnamefont {Jones}}, \bibinfo {author} {\bibfnamefont {P.~K.}\ \bibnamefont
  {Thakur}}, \bibinfo {author} {\bibfnamefont {T.-L.}\ \bibnamefont {Lee}},
  \bibinfo {author} {\bibfnamefont {C.}~\bibnamefont {Kalha}}, \bibinfo
  {author} {\bibfnamefont {C.}~\bibnamefont {Schlueter}}, \bibinfo {author}
  {\bibfnamefont {T.~D.}\ \bibnamefont {Veal}}, \bibinfo {author}
  {\bibfnamefont {J.~B.}\ \bibnamefont {Varley}}, \bibinfo {author}
  {\bibfnamefont {M.~R.}\ \bibnamefont {Wagner}},\ and\ \bibinfo {author}
  {\bibfnamefont {A.}~\bibnamefont {Regoutz}},\ }\href
  {https://doi.org/https://doi.org/10.1002/adma.202204217} {\bibfield
  {journal} {\bibinfo  {journal} {Adv. Mater.}\ }\textbf {\bibinfo {volume}
  {34}},\ \bibinfo {pages} {2204217} (\bibinfo {year} {2022})}\BibitemShut
  {NoStop}%
\bibitem [{\citenamefont {Anber}\ \emph {et~al.}(2020)\citenamefont {Anber},
  \citenamefont {Foley}, \citenamefont {Lang}, \citenamefont {Nathaniel},
  \citenamefont {Hart}, \citenamefont {Tadjer}, \citenamefont {Hobart},
  \citenamefont {Pearton},\ and\ \citenamefont {Taheri}}]{anber2020structural}%
  \BibitemOpen
  \bibfield  {author} {\bibinfo {author} {\bibfnamefont {E.~A.}\ \bibnamefont
  {Anber}}, \bibinfo {author} {\bibfnamefont {D.}~\bibnamefont {Foley}},
  \bibinfo {author} {\bibfnamefont {A.~C.}\ \bibnamefont {Lang}}, \bibinfo
  {author} {\bibfnamefont {J.}~\bibnamefont {Nathaniel}}, \bibinfo {author}
  {\bibfnamefont {J.~L.}\ \bibnamefont {Hart}}, \bibinfo {author}
  {\bibfnamefont {M.~J.}\ \bibnamefont {Tadjer}}, \bibinfo {author}
  {\bibfnamefont {K.~D.}\ \bibnamefont {Hobart}}, \bibinfo {author}
  {\bibfnamefont {S.}~\bibnamefont {Pearton}},\ and\ \bibinfo {author}
  {\bibfnamefont {M.~L.}\ \bibnamefont {Taheri}},\ }\href
  {https://doi.org/10.1063/5.0022170} {\bibfield  {journal} {\bibinfo
  {journal} {Appl. Phys. Lett.}\ }\textbf {\bibinfo {volume} {117}},\ \bibinfo
  {pages} {152101} (\bibinfo {year} {2020})}\BibitemShut {NoStop}%
\bibitem [{\citenamefont {Azarov}\ \emph {et~al.}(2023)\citenamefont {Azarov},
  \citenamefont {Fern{\'a}ndez}, \citenamefont {Zhao}, \citenamefont
  {Djurabekova}, \citenamefont {He}, \citenamefont {He}, \citenamefont {Prytz},
  \citenamefont {Vines}, \citenamefont {Bektas}, \citenamefont {Chekhonin},
  \citenamefont {Klingner}, \citenamefont {Hlawacek},\ and\ \citenamefont
  {Kuznetsov}}]{azarov2023universal}%
  \BibitemOpen
  \bibfield  {author} {\bibinfo {author} {\bibfnamefont {A.}~\bibnamefont
  {Azarov}}, \bibinfo {author} {\bibfnamefont {J.~G.}\ \bibnamefont
  {Fern{\'a}ndez}}, \bibinfo {author} {\bibfnamefont {J.}~\bibnamefont {Zhao}},
  \bibinfo {author} {\bibfnamefont {F.}~\bibnamefont {Djurabekova}}, \bibinfo
  {author} {\bibfnamefont {H.}~\bibnamefont {He}}, \bibinfo {author}
  {\bibfnamefont {R.}~\bibnamefont {He}}, \bibinfo {author} {\bibfnamefont
  {{\O}.}~\bibnamefont {Prytz}}, \bibinfo {author} {\bibfnamefont
  {L.}~\bibnamefont {Vines}}, \bibinfo {author} {\bibfnamefont
  {U.}~\bibnamefont {Bektas}}, \bibinfo {author} {\bibfnamefont
  {P.}~\bibnamefont {Chekhonin}}, \bibinfo {author} {\bibfnamefont
  {N.}~\bibnamefont {Klingner}}, \bibinfo {author} {\bibfnamefont
  {G.}~\bibnamefont {Hlawacek}},\ and\ \bibinfo {author} {\bibfnamefont
  {A.}~\bibnamefont {Kuznetsov}},\ }\bibfield  {journal} {\bibinfo  {journal}
  {arXiv preprint arXiv:2303.13114}\ }\href
  {https://doi.org/10.48550/arXiv.2303.13114} {10.48550/arXiv.2303.13114}
  (\bibinfo {year} {2023})\BibitemShut {NoStop}%
\bibitem [{\citenamefont {Yoshioka}\ \emph {et~al.}(2007)\citenamefont
  {Yoshioka}, \citenamefont {Hayashi}, \citenamefont {Kuwabara}, \citenamefont
  {Oba}, \citenamefont {Matsunaga},\ and\ \citenamefont
  {Tanaka}}]{yoshioka2007structures}%
  \BibitemOpen
  \bibfield  {author} {\bibinfo {author} {\bibfnamefont {S.}~\bibnamefont
  {Yoshioka}}, \bibinfo {author} {\bibfnamefont {H.}~\bibnamefont {Hayashi}},
  \bibinfo {author} {\bibfnamefont {A.}~\bibnamefont {Kuwabara}}, \bibinfo
  {author} {\bibfnamefont {F.}~\bibnamefont {Oba}}, \bibinfo {author}
  {\bibfnamefont {K.}~\bibnamefont {Matsunaga}},\ and\ \bibinfo {author}
  {\bibfnamefont {I.}~\bibnamefont {Tanaka}},\ }\href
  {https://doi.org/10.1088/0953-8984/19/34/346211} {\bibfield  {journal}
  {\bibinfo  {journal} {J. Phys.: Condens. Matter}\ }\textbf {\bibinfo {volume}
  {19}},\ \bibinfo {pages} {346211} (\bibinfo {year} {2007})}\BibitemShut
  {NoStop}%
\bibitem [{\citenamefont {Ma}\ \emph {et~al.}(2016)\citenamefont {Ma},
  \citenamefont {Tanen}, \citenamefont {Verma}, \citenamefont {Guo},
  \citenamefont {Luo}, \citenamefont {Xing},\ and\ \citenamefont
  {Jena}}]{ma2016intrinsic}%
  \BibitemOpen
  \bibfield  {author} {\bibinfo {author} {\bibfnamefont {N.}~\bibnamefont
  {Ma}}, \bibinfo {author} {\bibfnamefont {N.}~\bibnamefont {Tanen}}, \bibinfo
  {author} {\bibfnamefont {A.}~\bibnamefont {Verma}}, \bibinfo {author}
  {\bibfnamefont {Z.}~\bibnamefont {Guo}}, \bibinfo {author} {\bibfnamefont
  {T.}~\bibnamefont {Luo}}, \bibinfo {author} {\bibfnamefont {H.~G.}\
  \bibnamefont {Xing}},\ and\ \bibinfo {author} {\bibfnamefont
  {D.}~\bibnamefont {Jena}},\ }\href {https://doi.org/10.1063/1.4968550}
  {\bibfield  {journal} {\bibinfo  {journal} {Appl. Phys. Lett.}\ }\textbf
  {\bibinfo {volume} {109}},\ \bibinfo {pages} {212101} (\bibinfo {year}
  {2016})}\BibitemShut {NoStop}%
\bibitem [{\citenamefont {Dingwell}(1992)}]{dingwell1992density}%
  \BibitemOpen
  \bibfield  {author} {\bibinfo {author} {\bibfnamefont {D.~B.}\ \bibnamefont
  {Dingwell}},\ }\href
  {https://doi.org/https://doi.org/10.1111/j.1151-2916.1992.tb04239.x}
  {\bibfield  {journal} {\bibinfo  {journal} {J. Am. Ceram. Soc.}\ }\textbf
  {\bibinfo {volume} {75}},\ \bibinfo {pages} {1656} (\bibinfo {year}
  {1992})}\BibitemShut {NoStop}%
\bibitem [{\citenamefont {Yu}\ \emph {et~al.}(2003)\citenamefont {Yu},
  \citenamefont {Overgaard}, \citenamefont {Droopad}, \citenamefont
  {Passlack},\ and\ \citenamefont {Abrokwah}}]{yu2003growth}%
  \BibitemOpen
  \bibfield  {author} {\bibinfo {author} {\bibfnamefont {Z.}~\bibnamefont
  {Yu}}, \bibinfo {author} {\bibfnamefont {C.~D.}\ \bibnamefont {Overgaard}},
  \bibinfo {author} {\bibfnamefont {R.}~\bibnamefont {Droopad}}, \bibinfo
  {author} {\bibfnamefont {M.}~\bibnamefont {Passlack}},\ and\ \bibinfo
  {author} {\bibfnamefont {J.~K.}\ \bibnamefont {Abrokwah}},\ }\href
  {https://doi.org/10.1063/1.1572478} {\bibfield  {journal} {\bibinfo
  {journal} {Appl. Phys. Lett.}\ }\textbf {\bibinfo {volume} {82}},\ \bibinfo
  {pages} {2978} (\bibinfo {year} {2003})}\BibitemShut {NoStop}%
\bibitem [{\citenamefont {Galazka}\ \emph {et~al.}(2022)\citenamefont
  {Galazka}, \citenamefont {Ganschow}, \citenamefont {Seyidov}, \citenamefont
  {Irmscher}, \citenamefont {Pietsch}, \citenamefont {Chou}, \citenamefont
  {Bin~Anooz}, \citenamefont {Grueneberg}, \citenamefont {Popp}, \citenamefont
  {Dittmar}, \citenamefont {Kwasniewski}, \citenamefont {Suendermann},
  \citenamefont {Klimm}, \citenamefont {Straubinger}, \citenamefont
  {Schroeder},\ and\ \citenamefont {Bickermann}}]{galazka2022two}%
  \BibitemOpen
  \bibfield  {author} {\bibinfo {author} {\bibfnamefont {Z.}~\bibnamefont
  {Galazka}}, \bibinfo {author} {\bibfnamefont {S.}~\bibnamefont {Ganschow}},
  \bibinfo {author} {\bibfnamefont {P.}~\bibnamefont {Seyidov}}, \bibinfo
  {author} {\bibfnamefont {K.}~\bibnamefont {Irmscher}}, \bibinfo {author}
  {\bibfnamefont {M.}~\bibnamefont {Pietsch}}, \bibinfo {author} {\bibfnamefont
  {T.-S.}\ \bibnamefont {Chou}}, \bibinfo {author} {\bibfnamefont
  {S.}~\bibnamefont {Bin~Anooz}}, \bibinfo {author} {\bibfnamefont
  {R.}~\bibnamefont {Grueneberg}}, \bibinfo {author} {\bibfnamefont
  {A.}~\bibnamefont {Popp}}, \bibinfo {author} {\bibfnamefont {A.}~\bibnamefont
  {Dittmar}}, \bibinfo {author} {\bibfnamefont {A.}~\bibnamefont
  {Kwasniewski}}, \bibinfo {author} {\bibfnamefont {M.}~\bibnamefont
  {Suendermann}}, \bibinfo {author} {\bibfnamefont {D.}~\bibnamefont {Klimm}},
  \bibinfo {author} {\bibfnamefont {T.}~\bibnamefont {Straubinger}}, \bibinfo
  {author} {\bibfnamefont {T.}~\bibnamefont {Schroeder}},\ and\ \bibinfo
  {author} {\bibfnamefont {M.}~\bibnamefont {Bickermann}},\ }\href
  {https://doi.org/10.1063/5.0086996} {\bibfield  {journal} {\bibinfo
  {journal} {Appl. Phys. Lett.}\ }\textbf {\bibinfo {volume} {120}},\ \bibinfo
  {pages} {152101} (\bibinfo {year} {2022})}\BibitemShut {NoStop}%
\bibitem [{\citenamefont {Heinselman}\ \emph {et~al.}(2022)\citenamefont
  {Heinselman}, \citenamefont {Haven}, \citenamefont {Zakutayev},\ and\
  \citenamefont {Reese}}]{heinselman2022projected}%
  \BibitemOpen
  \bibfield  {author} {\bibinfo {author} {\bibfnamefont {K.~N.}\ \bibnamefont
  {Heinselman}}, \bibinfo {author} {\bibfnamefont {D.}~\bibnamefont {Haven}},
  \bibinfo {author} {\bibfnamefont {A.}~\bibnamefont {Zakutayev}},\ and\
  \bibinfo {author} {\bibfnamefont {S.~B.}\ \bibnamefont {Reese}},\ }\href
  {https://doi.org/10.1021/acs.cgd.2c00340} {\bibfield  {journal} {\bibinfo
  {journal} {Cryst. Growth Des.}\ }\textbf {\bibinfo {volume} {22}},\ \bibinfo
  {pages} {4854} (\bibinfo {year} {2022})}\BibitemShut {NoStop}%
\bibitem [{\citenamefont {Kresse}\ and\ \citenamefont
  {Hafner}(1993)}]{vasp1993}%
  \BibitemOpen
  \bibfield  {author} {\bibinfo {author} {\bibfnamefont {G.}~\bibnamefont
  {Kresse}}\ and\ \bibinfo {author} {\bibfnamefont {J.}~\bibnamefont
  {Hafner}},\ }\href {https://doi.org/10.1103/PhysRevB.47.558} {\bibfield
  {journal} {\bibinfo  {journal} {Phys. Rev. B}\ }\textbf {\bibinfo {volume}
  {47}},\ \bibinfo {pages} {558} (\bibinfo {year} {1993})}\BibitemShut
  {NoStop}%
\bibitem [{\citenamefont {Bl{\"o}chl}(1994)}]{paw1994}%
  \BibitemOpen
  \bibfield  {author} {\bibinfo {author} {\bibfnamefont {P.~E.}\ \bibnamefont
  {Bl{\"o}chl}},\ }\href {https://doi.org/10.1103/PhysRevB.50.17953} {\bibfield
   {journal} {\bibinfo  {journal} {Phys. Rev. B}\ }\textbf {\bibinfo {volume}
  {50}},\ \bibinfo {pages} {17953} (\bibinfo {year} {1994})}\BibitemShut
  {NoStop}%
\bibitem [{\citenamefont {Perdew}\ \emph {et~al.}(1996)\citenamefont {Perdew},
  \citenamefont {Burke},\ and\ \citenamefont {Ernzerhof}}]{pbe1996}%
  \BibitemOpen
  \bibfield  {author} {\bibinfo {author} {\bibfnamefont {J.~P.}\ \bibnamefont
  {Perdew}}, \bibinfo {author} {\bibfnamefont {K.}~\bibnamefont {Burke}},\ and\
  \bibinfo {author} {\bibfnamefont {M.}~\bibnamefont {Ernzerhof}},\ }\href
  {https://doi.org/10.1103/PhysRevLett.77.3865} {\bibfield  {journal} {\bibinfo
   {journal} {Phys. Rev. Lett.}\ }\textbf {\bibinfo {volume} {77}},\ \bibinfo
  {pages} {3865} (\bibinfo {year} {1996})}\BibitemShut {NoStop}%
\bibitem [{\citenamefont {Byggm{\"a}star}\ \emph
  {et~al.}(2022{\natexlab{b}})\citenamefont {Byggm{\"a}star}, \citenamefont
  {Nikoulis}, \citenamefont {Fellman}, \citenamefont {Granberg}, \citenamefont
  {Djurabekova},\ and\ \citenamefont {Nordlund}}]{byggmastar2022multiscale}%
  \BibitemOpen
  \bibfield  {author} {\bibinfo {author} {\bibfnamefont {J.}~\bibnamefont
  {Byggm{\"a}star}}, \bibinfo {author} {\bibfnamefont {G.}~\bibnamefont
  {Nikoulis}}, \bibinfo {author} {\bibfnamefont {A.}~\bibnamefont {Fellman}},
  \bibinfo {author} {\bibfnamefont {F.}~\bibnamefont {Granberg}}, \bibinfo
  {author} {\bibfnamefont {F.}~\bibnamefont {Djurabekova}},\ and\ \bibinfo
  {author} {\bibfnamefont {K.}~\bibnamefont {Nordlund}},\ }\href
  {https://doi.org/10.1088/1361-648x/ac6f39} {\bibfield  {journal} {\bibinfo
  {journal} {J. Phys.: Condens. Matter}\ }\textbf {\bibinfo {volume} {34}},\
  \bibinfo {pages} {305402} (\bibinfo {year} {2022}{\natexlab{b}})}\BibitemShut
  {NoStop}%
\bibitem [{qui()}]{quip}%
  \BibitemOpen
  \href {https://github.com/libAtoms/QUIP} {\bibinfo {title} {Quip--quantum
  mechanics and interatomic potentials,
  https://github.com/libatoms/quip}}\BibitemShut {NoStop}%
\bibitem [{\citenamefont {Togo}\ and\ \citenamefont
  {Tanaka}(2015)}]{phonopy2015}%
  \BibitemOpen
  \bibfield  {author} {\bibinfo {author} {\bibfnamefont {A.}~\bibnamefont
  {Togo}}\ and\ \bibinfo {author} {\bibfnamefont {I.}~\bibnamefont {Tanaka}},\
  }\href {https://doi.org/https://doi.org/10.1016/j.scriptamat.2015.07.021}
  {\bibfield  {journal} {\bibinfo  {journal} {Scr. Mater.}\ }\textbf {\bibinfo
  {volume} {108}},\ \bibinfo {pages} {1} (\bibinfo {year} {2015})}\BibitemShut
  {NoStop}%
\bibitem [{\citenamefont {{Le Roux}}\ and\ \citenamefont
  {Jund}(2010)}]{leroux2010ring}%
  \BibitemOpen
  \bibfield  {author} {\bibinfo {author} {\bibfnamefont {S.}~\bibnamefont {{Le
  Roux}}}\ and\ \bibinfo {author} {\bibfnamefont {P.}~\bibnamefont {Jund}},\
  }\href {https://doi.org/10.1016/j.commatsci.2010.04.023} {\bibfield
  {journal} {\bibinfo  {journal} {Comput. Mater. Sci.}\ }\textbf {\bibinfo
  {volume} {49}},\ \bibinfo {pages} {70} (\bibinfo {year} {2010})}\BibitemShut
  {NoStop}%
\end{thebibliography}%

\section*{Acknowledgements}

This work is supported by Guangdong Basic and Applied Basic Research Foundation under Grant 2023A1515012048 and Shenzhen Fundamental Research Program under Grant JCYJ20220530114615035.
F.D. acknowledges M-ERA.NET Program for financial support via GOFIB project administrated in Finland by the Academy of Finland project number 352518. 
F.D., K.N. and J.B. acknowledge the international collaboration within the COST Action FIT4NANO CA19140 supported by the European Cooperation in Science and Technology, \url{https://www.cost.eu/}.
The computational resource is supported by the Center for Computational Science and Engineering at the Southern University of Science and Technology.
The authors are also grateful to the grants of computer power from CSC – IT Center for Science, Finland. 
J.Z. also acknowledge Prof. L.-J. Zhang and Prof. Y.-H. Fu at Jilin University, for the helpful discussion on the CALYPSO-predicted $P\overline{1}$ and $Pmc2_{1}$ \ce{Ga2O3} structures~\cite{wang2020discovery}. 
The authors are also grateful to Dr. H. Liu and Dr. I. Makkonen at the University of Helsinki and Prof. A. Kuznetsov at the University of Oslo for the insightful discussion.  

\section*{Author contributions}

J.Z. initiated and coordinated the study.
J.Z. developed the core DFT database and fitted the initial soapGAP version of the interatomic potentials;
J.B. developed the concept of the tabGAP version of the interatomic potentials and composed the code package.
J.B. and H.H. fitted the initial tabGAP version of the interatomic potentials; 
J.Z., J.B., and H.H. performed iterative optimization of both interatomic potentials and analyzed the validation, reliability and performance; 
J.Z. conducted and analyzed the MD simulations shown in this work;
J.Z. composed the first draft of the paper with the input from J.B., F.D., and H.H.;
F.D., K.N., and M.H. revised the final version of the paper; 
All authors contributed to the scientific discussion throughout the whole work.

\section*{Competing interests}

The authors declare no competing interests.

\section*{Additional information}

\textbf{Supplementary information} is available for this paper at \url{https://doi.org/...}

\textbf{Correspondence} and requests for materials should be addressed to J.Z. (Email: \url{zhaojl@sustech.edu.cn}) and M.H. (Email: \url{huamy@sustech.edu.cn}).

\end{document}